\documentclass[12pt]{article}

\usepackage{newtxtext,newtxmath}
\usepackage[T1]{fontenc}

\usepackage{graphicx}

\usepackage[letterpaper,margin=1in]{geometry}

\renewenvironment{abstract}
	{\quotation}
	{\endquotation}

\date{}

\makeatletter
\renewcommand{\fnum@figure}{\textbf{Figure \thefigure}}
\renewcommand{\fnum@table}{\textbf{Table \thetable}}
\makeatother

\usepackage{scicite}

\usepackage{url}

\def\scititle{
	Direct Asymmetric Photochemistry with Synthetic Chiral Light
}
\title{\bfseries \boldmath \scititle}

\author{
	Jason B. Greenwood$^{1\ast}$, Letizia Fede$^{2}$, Daniel García González$^{3}$,\and 
	Umberto Dellasette$^{2}$, Ross Anderson$^{4}$, Chris Sparling$^{4}$, Dave Townsend$^{4}$,\and
    Patricia Vindel-Zandbergen$^{5}$, David Ayuso$^{6}$, Val\'erie Blanchet$^{2}$\and 
    Bernard Pons$^{2}$, Andr\'es Ord\'o\~nez$^{3\ast}$,
	Yann Mairesse$^{2\ast}$\and
    \small$^{1}$Centre for Light-Matter Interactions, School of Mathematics and Physics, Queen’s University Belfast, Belfast, UK.\and\
	\small$^{2}$Universit\'e de Bordeaux -- CNRS -- CEA, CELIA, UMR5107, Talence, France.\and\
	\small$^{3}$Department of Physics, Freie Universit\"at, 14195 Berlin, Germany.\and\
	\small$^{4}$Institute of Photonics and Quantum Sciences, Heriot-Watt University, Edinburgh EH14 4AS, UK.\and\
    \small$^{5}$Department of Chemistry, New York University, New York 10003, USA.\and\
    \small$^{6}$Department of Chemistry, Imperial College London, W12 0BZ London, UK.\and
	\small$^\ast$Corresponding author. Email:  j.greenwood@qub.ac.uk (J.G.);\\ \small andres.ordonez@fu-berlin.de (A.O.); yann.mairesse@u-bordeaux.fr (Y.M.)\and
}

\begin{document} 

\maketitle

\begin{abstract} \bfseries \boldmath

Circularly polarized light has long been used to distinguish the two mirror images of chiral molecules. Recently, a conceptual breakthrough has predicted that a new form of optical fields - synthetic chiral light - could usher in an \textit{`electric–dipole revolution'} in chiroptical interactions. We report the first experimental evidence for this prediction in the optical domain, using a three-dimensional laser electric field synthesized to trace a chiral pattern. The ionization and fragmentation of chiral molecules was found to depend on both the molecular handedness, and the handedness and geometry of the synthetic chiral field.  Supported by two theoretical models, our results demonstrate that synthetic chiral light can induce strong chiroptical interactions, opening new horizons for asymmetric photochemistry and chiral analysis.

\end{abstract}

\begin{figure} 
	\centering
	\includegraphics[width=0.8\textwidth]{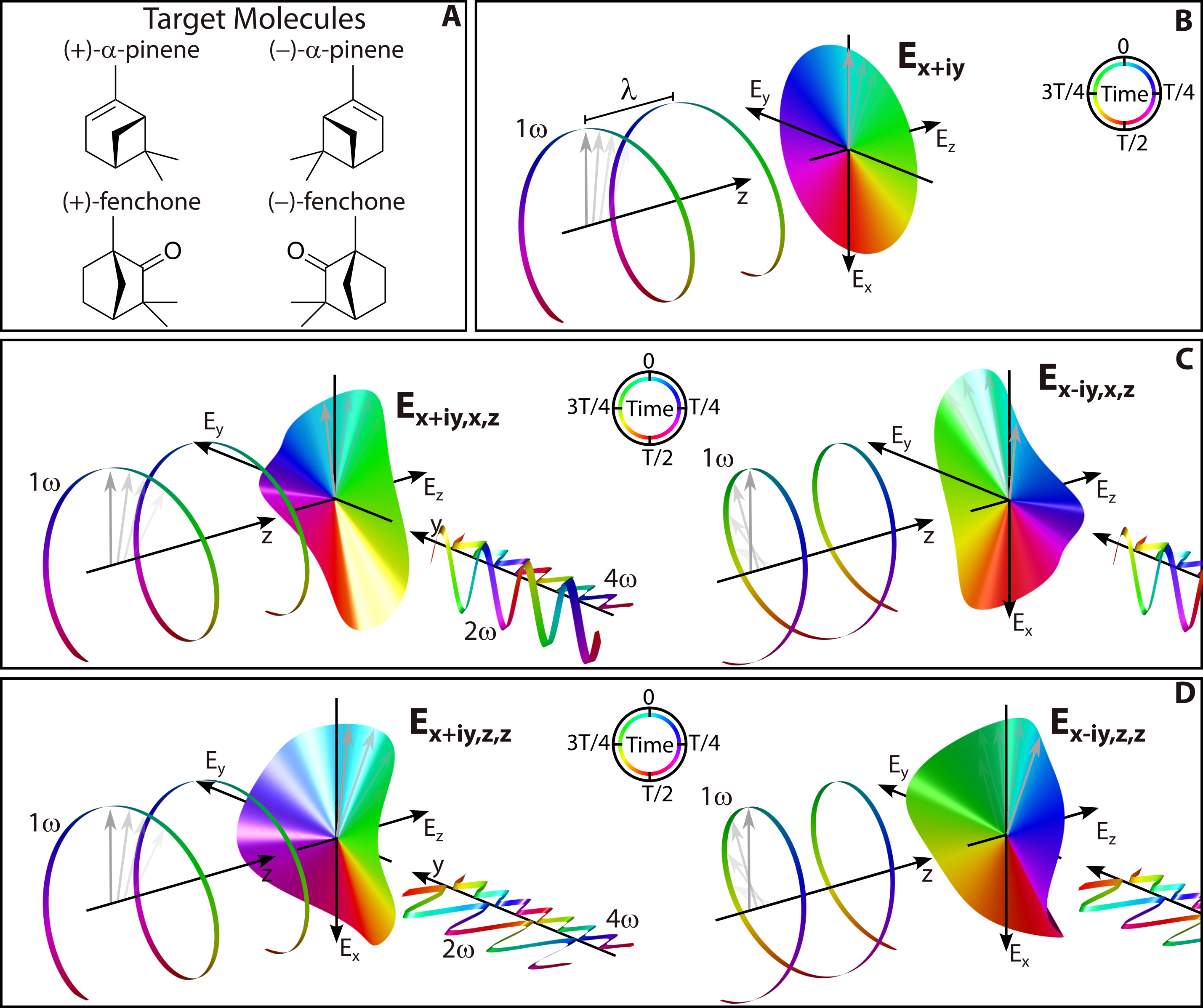} 
	\caption{\textbf{Generation of synthetic chiral light.}
		\textbf{(A)} Target molecules used in the experiment: $(+)$ and $(-)$ enantiomers of  $\alpha$-pinene (top), and fenchone (bottom). \textbf{(B)} The electric field of circularly polarized light traces out a helix in space as it propagates along $z$ but its pitch $\lambda$ is a thousand times larger than the typical size of a chiral molecule. This means on the molecular scale, within the electric-dipole approximation, the electric field oscillates only in a plane ($E_x,E_y$), generally making chiroptical processes very weak. \textbf{(C)} A circularly polarized fundamental laser beam ($1\omega$ with oscillation period T) is intersected with collinear second and fourth harmonics ($2\omega$ and $4\omega$) propagating in a direction orthogonal to the $1\omega$ beam. With the $2\omega$ and $4\omega$ pulses linearly polarized in perpendicular directions, an electric field ($E_{x\pm iy,x,z}$) oscillating in three dimensions is produced whose Lissajous figure forms a chiral pattern, enabling strong enantio-sensitive coupling of the field to chiral molecules. The chirality of the synthetic light field can be reversed by switching the helicity of the $1\omega$ circular polarization, which is equivalent to a reflection in the $xz$ plane (right panel). \textbf{(D)} A different synthetic chiral light configuration ($E_{x\pm iy,z,z}$) obtained with the $2\omega$ and $4\omega$ polarizations parallel. In all panels, the colormap encodes the time, T$\approx 3.8$~fs being the period of the $1\omega$ field.}
	\label{fig1} 
\end{figure}

\noindent
Chiral enantiomers are pairs of molecules which, like our hands, are mirror images that cannot be superimposed by rotations alone (Fig. \ref{fig1}A). As a result, the interaction of each enantiomer with another chiral object can be profoundly different. This is critical in biological systems, where the actions of enzymes, hormones, and pharmaceuticals can depend strongly on ‘handshaking’ with acceptor sites.
We are also on the cusp of a revolution in diagnostic healthcare where the presence of small amounts of non-innate enantiomers is emerging as a key indicator of aging and the pathology of diseases such as cancer, Alzheimer’s, and diabetes \cite{liu2023a}.

Enantiomers have identical physical properties and can only be discriminated through interactions with another chiral object, such as in an asymmetric chemical reaction or the differential drag in a chiral chromatography column. Circularly polarized light exhibits handedness-dependent absorption, making circular dichroism (CD) a powerful tool for chiral analysis \cite{berova2000}.  This effect relies on the sensitivity of the molecular response to the chiral path mapped out by circularly polarized light as it propagates – a left- or right-handed helix (Fig. \ref{fig1}B). In terms of light-matter interactions, this helicoidal structure is only felt by the molecule beyond the electric-dipole approximation. The pitch of the helix is the wavelength $\lambda$ of the light, which is several hundred nanometres in the visible range, 3 orders of magnitude larger than small to medium-sized chiral molecules. This limits the differential strength of the chiral interactions and hence the enantio-sensitivity of optical techniques or their ability to manipulate chiral matter.

Much stronger chiral light-matter interactions could be realized through enantio-dependent electric-dipole coupling. Within the electric-dipole approximation, the electric field of circularly polarized light only draws a circle in time (Fig. \ref{fig1}B), a non-chiral structure which cannot discriminate enantiomers, but stronger enantio-sensitivity can be achieved if a `chiral observer' is employed \cite{ordonez2018, ayuso2022}. For instance, by resolving the angular distribution of photoionized electrons in photoelectron circular dichroism \cite{ritchie1975,bowering2001,nahon2015,sparling2025}, or of the ionic fragments in Coulomb explosion imaging \cite{pitzer2013}. However, while these methods enable efficient chiral sensing through angularly resolved measurements\cite{comby2018a}, the total photoemission or fragmentation yields are not enantio-sensitive (within the electric-dipole approximation), posing a fundamental challenge for direct asymmetric photochemistry. Achieving high enantio-sensitivity at the level of total yields requires an efficient chiral photonic reagent: synthetic chiral light.

\subsection*{Synthetic chiral light}
 Strong enantiosensitivity should become possible for overall product yields if the light’s electric field oscillations map out a chiral shape \cite{ayuso2019}.  The three-dimensional field of this synthetic chiral light can be generated by crossing laser beams of different frequencies and polarizations, enabling enantio-dependent electric-dipole interactions where the beams overlap (Fig. \ref{fig1}C,D) \cite{ayuso2019,neufeld2021,ayuso2022,ye2023,habibovic2024,ordonez2025a,smirnova2025,kohnke2026}. These strong chiroptical interactions could lead to much more sensitive chiral sensing techniques \cite{smirnova2025} and enable enantio-selective photolysis\cite{moore2024}, induction of chirality in achiral systems\cite{romao2025,zeng2025}, or development of chiroptical tweezers \cite{cameron2023}.

In the low energy range of the electromagnetic spectrum, the combination of three microwave fields with mutually orthogonal polarizations was theoretically predicted \cite{frishman2003,leibscher2019} and experimentally shown \cite{eibenberger2017,perez2017,lee2022} to produce resonant rotational excitation with very high enantio-sensitivity. This can be seen as a realization of synthetic chiral microwave radiation. In the optical domain however, where the use of ultrashort laser pulses opens unique perspectives for measurement and control, the \textit{`electric–dipole revolution'} in chiral photochemistry \cite{ayuso2022,smirnova2025} had not been experimentally realized until now. This could be rationalized by the practical challenges involved, such as achieving spatial and temporal overlap of multiple non-collinear ultrashort laser pulses, minimizing the contribution of parasitic non-chiral pathways, and maintaining the chiral interaction as the relative phases of the beams change across the intersection volume,  a requirement known as global chirality\cite{ayuso2019,ayuso2022,rego2023,ye2023,ordonez2025a}. The generation of a 3D polarized field was very recently reported  by combining two non-collinear laser pulses \cite{kohnke2026} to excite a superposition of states in an atom which are not accessible with conventional light fields. However, to perform enantio-dependent photochemistry, chiral molecules need to be irradiated by a synthetic field with global chiral characteristics. 

Here we produce globally chiral synthetic light in the optical domain by combining three pulses of different colors, as shown in Fig. \ref{fig1}C,D, and use it to achieve enantio-dependent photoionization and photolysis. A circularly polarized fundamental beam (1$\omega$ - 1030 nm) propagating along the $z$ direction was intersected perpendicularly with linearly polarized, 2nd (2$\omega$ - 515 nm) and 4th (4$\omega$ - 258 nm) harmonics both propagating along $y$. In the first chiral light configuration (Fig. \ref{fig1}C), the 2$\omega$ and 4$\omega$ polarization directions are mutually perpendicular while in the second configuration (Fig. \ref{fig1}D) they are parallel and along the 1$\omega$ propagation direction. These are labelled according to the polarization of the $1\omega$, $2\omega$ and $4\omega$ components of the field $E_{x\pm iy,x,z}$ and $E_{x\pm iy,z,z}$. 

In the $E_{x\pm iy,z,z}$ configuration (Fig. \ref{fig1}D), the relative phase between the 1$\omega$ and 2$\omega$/4$\omega$ beams changes across their interaction areas, causing the 3D chiral field to rotate around the propagation axis $z$ of the fundamental beam, but crucially its shape -- and chirality --  remain unchanged \cite{ye2023}. For the $E_{x\pm iy,x,z}$ configuration (Fig. \ref{fig1}C), when this phase changes there is a change in the shape of the chiral field but its handedness is preserved, as can be verified using a chiral correlation function \cite{ordonez2025a} (see Methods). Hence, for both these synthetic chiral light configurations the electric field is globally chiral, enabling us to observe enantio-dependent ionization yields integrated over the interaction volume. Mirror images of these light fields can be produced by changing the phase $\Phi_{4\omega}$ of the 4$\omega$ beam with respect to the  2$\omega$ beam by $\pi$, or equally by reversing the helicity of the 1$\omega$ beam. 

\subsection*{Measurement of synthetic chiral dichroism}
\begin{figure} 
	\centering
	\includegraphics[width=0.9\textwidth]{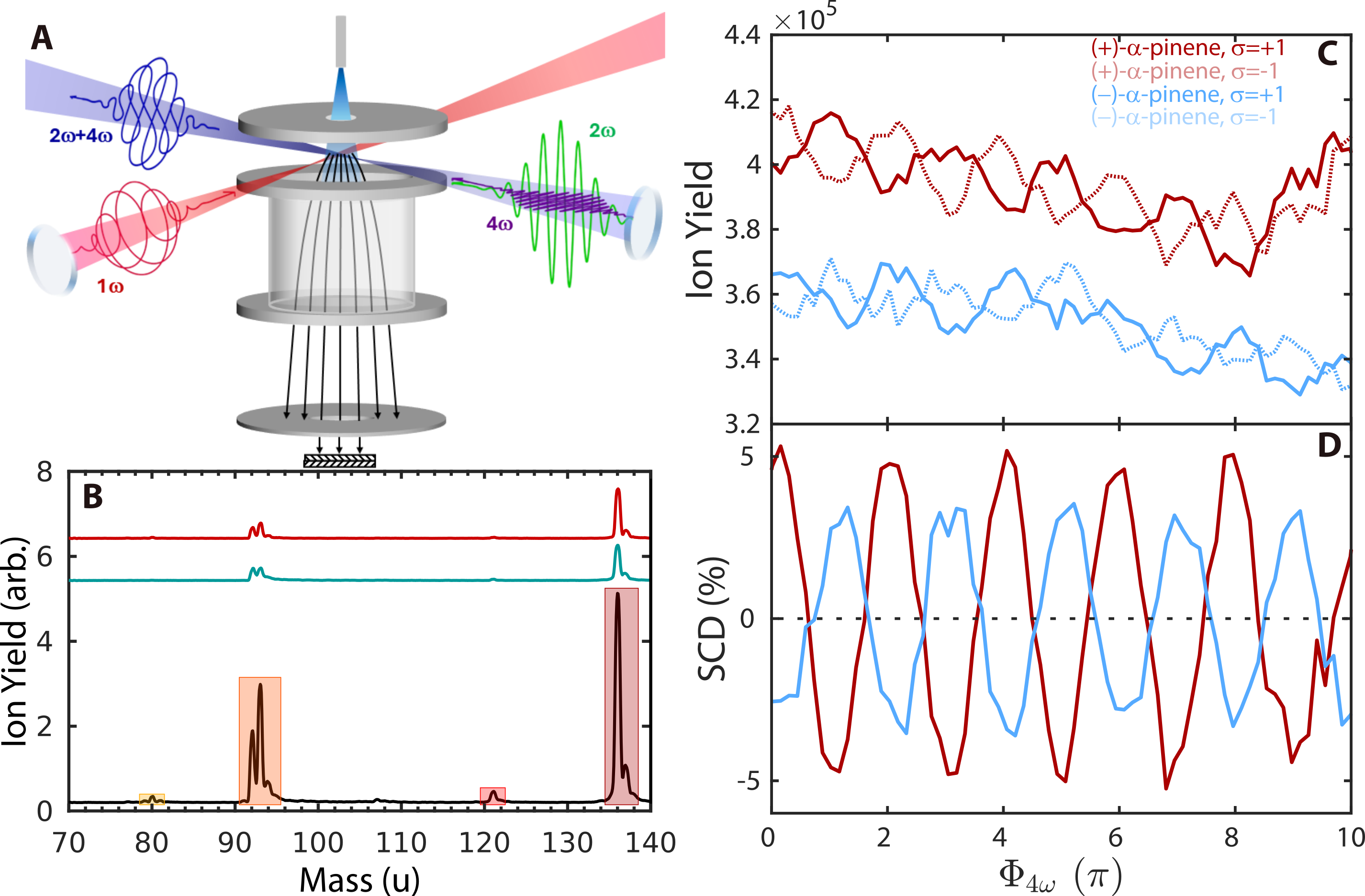} 
	\caption{\textbf{Measurement of Synthetic Chiral Dichroism with the $E_{x\pm iy,x,z}$ configuration in $\alpha$-pinene.}
    \textbf{(A)} The chiral time-of-flight (ToF) mass spectrometer. The orthogonally propagating 1$\omega$ and 2$\omega$/4$\omega$ pulses intersect an effusive gas jet on the axis of the mass spectrometer. An electric field extracts the resulting ions into a field free drift region where ions of different mass temporally separate before hitting a micro-channel plate detector. \textbf{(B)} The mass spectrum of $\alpha$-pinene obtained with $1\omega$ only (top-red), $2\omega$+$4\omega$ (middle-blue) and $1\omega$+$2\omega$+$4\omega$ (bottom-black). The plots have been shifted vertically for clarity. The shaded regions indicate the time gates, shown on the mass axis, used to measure yields of the parent molecular ion (mass 136) and fragment ions. \textbf{(C)} The raw parent molecular ion yields from synthetic chiral light interacting with (+)- or (-)-$\alpha$-pinene (red and blue curves), and with the  $1\omega$ component left- and right-circularly polarized $S_L$ (dark lines, $\sigma=+1$) and $S_R$ (lighter dashed lines, $\sigma=-1$). All signals clearly oscillate with the phase $\Phi_{4\omega}$ between the $2\omega$ and $4\omega$ fields (the four curves have been vertically offset to improve the visibility of the oscillations). The signals measured with opposite circular polarizations of the $1\omega$ field have a $\pi$ phase shift, as do the signals from the opposite enantiomers. \textbf{(D)} The Synthetic Chiral Dichroism (SCD) obtained by calculating the normalized difference of the yields obtained with left- and right-handed polarizations of the $1\omega$ field as a function of the $\Phi_{4\omega}$ phase.
		}
	\label{fig2} 
\end{figure}

We investigated the photoionization of fenchone and $\alpha$-pinene, two small chiral molecules with two stereogenic centers, whose rigid bicyclic structure ensures that only one conformer exists (Fig. \ref{fig1}(A)). The molecules flowed through an effusive jet into a time-of-flight (ToF) mass spectrometer, where they interacted with two crossed laser beams (Fig. \ref{fig2}A). The peak intensities of each laser pulse were, $I_\omega\sim10^{12}$~W.cm$^{-2}$,  $I_{2\omega}\sim6\times10^{10}$~W.cm$^{-2}$, $I_{4\omega}\sim6\times10^{8}$~W.cm$^{-2}$. In our optical setup (Fig. \ref{fig:optics}), the $\Phi_{4\omega}$ phase was controlled by the thickness of fused silica wedges present in the 2$\omega$/4$\omega$ beamline, and the circular polarization of 1$\omega$ reversed by a waveplate. The $\Phi_{4\omega}$ phase was calibrated relative to the wedge position using a stereo-photoelectron detector (see Methods).  

 Ions generated from the overlapping laser pulses were extracted into the ToF with the parent molecular ion and major fragment ions detected in a pulse counting mode (Figs. \ref{fig2}A, \ref{fig:TOF}). The spatial and temporal overlap was maximized by adjusting the beam positions and the delay of the 1$\omega$ pulse to achieve the highest count rates from ionization involving all 3 laser colors (Fig. \ref{fig:delay}). Even when the 3-pulse overlap was optimized, only a fraction of this yield originated from the 3-color ionization process, since the temporal overlap between the orthogonally propagating femtosecond laser pulses was only achieved in a fraction of the interaction volume (Fig. \ref{fig:crossco}). These background contributions, due to single pulses and a combination of 2 of the 3 colors, were measured for each ion $i$ to determine the fraction $F^i$ of the counts arising from ionization involving all 3 colors (see Methods and Table \ref{tab:Fraction F}).

The ion yield at each wedge position was obtained with the 1$\omega$ beam left- and right-circularly polarized and the background contribution to the average yield, (1-$F^i$), subtracted to extract the signal originating purely from 3-color ionization processes for each polarization state ($S_L^i$ and $S_R^i$). In Fig. \ref{fig2}C, $S_L$ and $S_R$ for the parent molecular ion have been plotted for both enantiomers of $\alpha$-pinene as a function of  $\Phi_{4\omega}$. Oscillations in the yield were clearly evident and underwent a $\pi$ phase change when the helicity of the 1$\omega$ field or the enantiomer was switched. The magnitude of this synthetic chiral dichroism (SCD) was quantified as the normalized difference of the yields. 
\begin{equation}
\textrm{SCD}^i=2\frac{S_L^i-S_R^i}{S_L^i+S_R^i}
\end{equation}
The measured SCD, plotted in Fig. \ref{fig2}D, shows periodic oscillations with an amplitude of up to 5\%  in (+)-$\alpha$-pinene. This is more than an order of magnitude higher than conventional one-color photoionization circular dichroism using either 2$\omega$ or 4$\omega$ only (see Table \ref{tab:PICD}). 
The SCD obtained in the opposite (-) enantiomer shows weaker modulations due to reduced stability of the $\Phi_{4\omega}$ phase during that measurement, resulting in a lower average for the SCD amplitude. 

\subsection*{Asymmetric photochemistry}
\begin{figure} 
	\centering
	\includegraphics[width=0.9\textwidth]{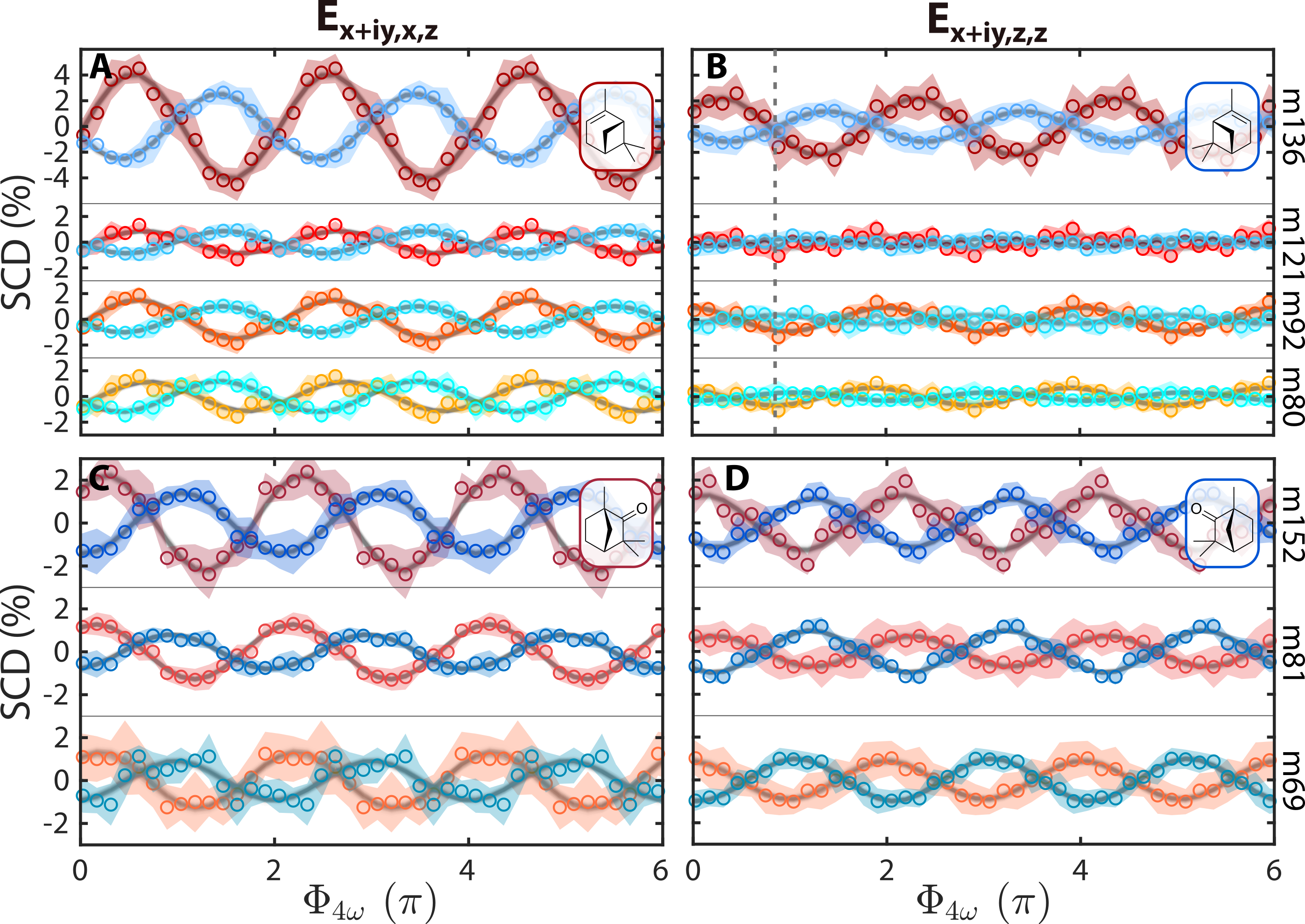} 
	\caption{\textbf{Enantio-sensitive photoionization and photolysis of $\alpha$-pinene and fenchone.}
   \textbf{(A-B)} Measured synthetic chiral dichroism in (+)-$\alpha$-pinene (red to orange) and ($-$)-$\alpha$-pinene (blue to cyan) as a function of the shape of the synthetic chiral light field $E_{x+iy,x,z}$ (A) and $E_{x+iy,z,z}$ (B), controlled by scanning the phase of the fourth harmonic beam with respect to the second harmonic beam. The different lines correspond to different ions, with m136 (parent), and m121, m92, m80 (fragments) from top to bottom. The colored shaded areas represent the 90\% confidence interval of the data, while 200 semi-transparent gray lines each represent a sinusoidal fit of randomly selected data sets within this confidence interval. The vertical dashed line in (B) corresponds to the phase at which the SCD of the parent is zero. \textbf{(C-D)}  Same measurements in (+)-fenchone (red to orange) and ($-$)-fenchone (dark to light blue). The different lines correspond to ions with m152 (parent), m81 and m69 (fragments) from top to bottom.\\
		}
	\label{fig3} 
\end{figure}

To improve the signal-to-noise ratio of the ionic fragments, which show weaker yields, we averaged the SCDs obtained in 6 consecutive periods and reconstructed this mean over a few oscillation cycles. We computed the 90\% confidence interval of this average by analyzing the dispersion of the 6 sets of data (see Methods). The values of the SCD obtained at $\pi$-shifted phases were also subtracted to anti-symmetrize the signal: 
\begin{equation}
\textrm{SCD}(\Phi_{4\omega})=\frac{1}{2}\left[\textrm{SCD}(\Phi_{4\omega})-\textrm{SCD}(\Phi_{4\omega}+\pi)\right]
\end{equation}
 Fig. \ref{fig3}A shows the results of this analysis for the parent and the most significant fragment ions of the enantiomers of $\alpha$-pinene.  

The SCD of the fragment ions, like the parent ion, exhibit oscillations with $\Phi_{4\omega}$, although with lower amplitudes. We performed sinusoidal fittings of 200 data sets randomly picked within the confidence interval of the measured points, and extracted the mean and confidence interval of the oscillation amplitude and phase. The gray shading in Fig. \ref{fig3} depicts these 200 fits as semi-transparent curves. The fitting parameters, presented in Table \ref{tab:SCDpinene}, show that the ions from opposite enantiomers have a $1.1\pm0.1\pi$ phase shift, confirming the genuine enantio-sensitivity of the signal. The amplitudes of the SCD modulations for (+)-$\alpha$-pinene ions with masses from 136, 121, 92/93 and 80, are 4.1, 0.9, 1.5 and 1.1\% respectively. All ions oscillate in phase, but the different amplitudes of their modulation imply that the branching ratios, defined as the yield of a given ion with respect to the total ionization yield, are modulated as a function of $\Phi_{4\omega}$. As shown in Fig. \ref{figBRD}, the value of $\Phi_{4\omega}$ minimizing the production of the parent ion in one enantiomer maximizes the relative contribution of the fragments of this enantiomer. This demonstrates that while the overall ionization yield can be maximized or minimized by adjusting the shape of the synthetic chiral light through control of the $\Phi_{4\omega}$ phase, selective dissociative ionization is also achievable.

To evaluate the sensitivity of SCD to the shape of the synthetic chiral light, we repeated the measurements in the  $E_{x+iy,z,z}$  configuration described in  Fig. \ref{fig1}D. Fig. \ref{fig3}B shows the resulting SCD which again has a phase reversal when the enantiomer is switched. In this configuration the 3-color contribution to the signal (fraction $F$, see Table \ref{tab:Fraction F}) is weaker, as is the overall SCD (ranging from 2.2\% for mass 136 down to 0.7\% for mass 80 in (+)-$\alpha$-pinene). Interestingly, the oscillations in the SCD of the parent ions and the lightest fragments are substantially shifted by 1.4$\pm0.4$ rad, i.e. around $\pi/2$. In Fig. \ref{fig3}B, this is visible at $\Phi_{4\omega}\approx \pi$: the synthetic chiral light produces the same amount of parent ions from both enantiomers, but maximizes the production of fragment ions with masses 80 and 92,93 in one enantiomer while minimizing it with its mirror image. This illustrates how synthetic chiral light can control the amplitudes and phases of different quantum pathways, thus enabling enantio-sensitive photochemistry.

\begin{figure} 
	\centering
	\includegraphics[width=0.9\textwidth]{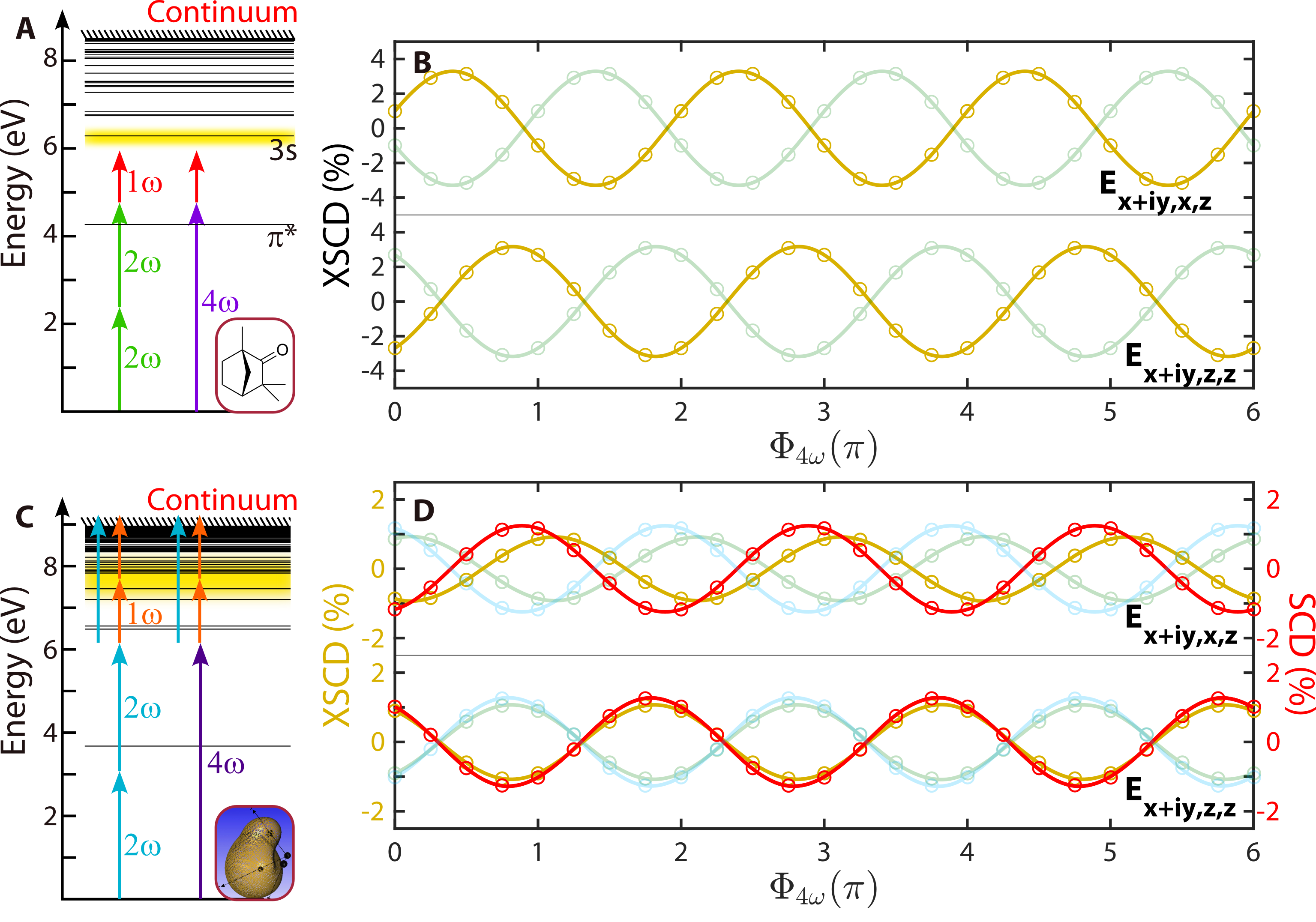} 
	\caption{\textbf{Theoretical calculations of Synthetic Chiral Dichroism.}
		\textbf{(A)} A key photoexcitation/ionization pathway in fenchone. The 3s state is excited by interference between pathway 1 (2$\omega$ + 2$\omega$ + 1$\omega$) and pathway 2 (4$\omega$ + 1$\omega$) leading to an enantiomer-dependent population rate. \textbf{(B)} Time-Dependent Schr\"odinger Equation (TDSE) calculations in fenchone of enantio-sensitive excitation synthetic chiral dichroism (XSCD) to the 3s state of (+) (yellow) and ($-$) (light green) fenchone using the configurations employed in the experiments  $E_{x+iy,x,z}$ and $E_{x+iy,z,z}$ as a function of the $\Phi_{4\omega}$ phase. \textbf{(C)} Energy levels and photoexcitation/ionization pathways in the 4-center model molecule. The isosurface in the inset displays the fundamental state of the molecule. \textbf{(D)} TDSE calculations of the XSCD (yellow and light green in opposite enantiomers) and SCD (red and light blue in opposite enantiomers) obtained for the 4-center model molecule, for the two chiral light configurations.
        }
	\label{fig4} 
\end{figure}

The results obtained in fenchone molecules are depicted in Fig. \ref{fig3}(C-D). They show clear oscillations of the SCD, with a (1.2$\pm$0.1)$\pi$ phase shift between enantiomers in the $E_{x+iy,x,z}$ configuration, and a (0.9$\pm$0.1)$\pi$ shift in the $E_{x+iy,z,z}$ configuration. The weaker SCD in (-)-fenchone is assigned to the 16\% lower enantiomeric excess of the sample compared to (+)-fenchone (Table \ref{tab:samples}). The maximum SCD, observed in the parent ion signal (m152), reaches 2.2\% in the $E_{x+iy,x,z}$ configuration and 1.3\% in the $E_{x+iy,z,z}$ configuration, a factor $\sim1.8$ lower than the value measured in $\alpha$-pinene. By contrast, the fragments show SCD values similar to, or even higher than $\alpha$-pinene fragment ions, although fenchone is less prone to dissociative ionization (Fig. \ref{fig:TOF}). The phase of the SCD oscillations show a statistically significant shift of around 0.3 rad relative to the parent ion yield for mass 81 in the $E_{x+iy,x,z}$ configuration, and around 0.5 rad for mass 69 in the $E_{x+iy,z,z}$ configuration. This is an additional demonstration of enantio-sensitive photolysis which can be controlled by the shape of the synthetic chiral light field. 

\subsection*{Modeling synthetic chiral excitation and ionization}
To support these measurements, we used two complementary theoretical approaches. First, we investigated the dichroism in the photoexcitation of neutral molecules by the synthetic chiral light. We solved the electronic time-dependent Schr\"odinger Equation (TDSE) for fenchone at its equilibrium geometry on a basis of unperturbed bound energy eigenstates, including all electric-dipole couplings, using the experimental intensities and polarizations, and averaging final populations over molecular orientations (see Methods). Our simulations show that the 3s state (Fig. \ref{fig4}A) is the most strongly populated photoexcited state (Fig. \ref{fig:SI_populations_perp}), as expected for the photon energies considered. To reveal the influence of the handedness of the synthetic chiral light on this population, we calculated the excitation synthetic chiral dichroism (XSCD), by comparing the populations $P_{L,R}$  obtained with left- and right-handed circularly polarized fundamental field: 
$\mathrm{XSCD}=2(P_L - P_R)/(P_L + P_R)$. The results (Fig. \ref{fig4}B) show enantiosensitive oscillations as a function of $\Phi_{4\omega}$, as observed in the measured SCDs. The XSCD reaches $\sim3\%$ for both chiral field configurations used in the experiments, but with a different phase dependence. More detailed analysis in Methods shows that these field configurations are represented by different chiral correlation functions, thereby accessing different properties of the molecule. Although the high intensities lead to many possible excitation pathways into 3s, the oscillations in Fig. \ref{fig4}B can be understood, in the simplest case, as resulting from the interference between the two pathways shown in Fig. \ref{fig4}A \cite{ordonez2025a} 
(see Methods). Higher excited states exhibit lower populations but larger XSCD values, which can change sign from state to state (see Fig. \ref{fig:SI_populations_perp}). 

To quantify how XSCD in excited state populations is transformed into SCD in the photoionization yield via absorption of additional photons, we turn to our second theoretical approach, which explicitly accounts for ionization. We performed TDSE calculations in a simple model chiral molecule which consists of an electron moving in the field of four nuclear centers. Nuclear charges and internuclear distances were adjusted to obtain an ionization potential similar to that of fenchone \cite{rozen2019}. Despite its simplistic structure, this 4-center molecule has been found to provide enantio-sensitive photoelectron angular distributions very similar to those measured in fenchone in a number of previous studies involving photoionization by multi-color light fields \cite{rozen2019,bloch2021,beaulieu2024}. Here we subject this molecule to synthetic chiral light composed of a fundamental $1\omega$ field at 800 nm wavelength and solve the TDSE including a discretized representation of the continuum (see Methods). The spectral width of the $1\omega$ pulse leads to substantial population of excited states in the $[7.00-8.25]$ eV range by $(2\times 2\omega+1\omega)$ and $(4\omega+1\omega)$ photon absorptions  (see Fig. \ref{fig4}C and Methods). The XSCD computed for this excitation manifold is presented in Fig. \ref{fig4}D for the $E_{x\pm iy,x,z}$ and $E_{x\pm iy,z,z}$ field configurations. In both cases, the XSCD maximizes around $1 \%$, which is somewhat lower than in the fenchone calculations, because the sum over individual excitation states with opposite XSCD signs reduces the signal (see Methods). However, both calculations show a substantial phase shift in the XSCD between the two field configurations.

The SCD resulting from subsequent ionization is presented in Fig. \ref{fig4}D. 
It shows the dichroism in the population of the continuum is slightly enhanced compared to the excited state, reaching around $1.2\%$ in both $E_{x\pm iy,x,z}$ and $E_{x\pm iy,z,z}$ field configurations.  Interestingly, in the $E_{x\pm iy,x,z}$ configuration, the SCD shows a clear phase shift with respect to the XSCD. This demonstrates the influence of the interference between multiple ionization pathways from the excited states, as displayed in Fig. \ref{fig4}C. Our calculations (Methods) show that the phase of the SCD oscillations, and therefore this interference, is very sensitive to the intensity of the 1$\omega$ field.

\subsection*{Conclusions}
Our measurements have demonstrated that synthetic chiral light, generated by intersecting laser beams of three distinct colors, can enantio-selectively ionize two different chiral molecules. The resulting differential ionization can be controlled on a sub-cycle timescale by varying the relative phase between two of the three beams, thereby modifying the three-dimensional electric-field trajectory which controls the handedness of the chiral light field and hence the enantio-sensitivity of the interaction. Chiral dichroism was found for both 3D light fields we synthesized and both molecules studied, but with different amplitude and phase dependencies. This demonstrates the potential for light fields to be shaped in such a way as to probe different quantum properties and pathways in molecules. 

Our theoretical simulations for fenchone show that this synthetic chiral dichroism originates from population asymmetries in excited states driven by interference between chiral and achiral excitation pathways involving all three photon energies. Calculations with the simple 4-center molecular model indicate that these asymmetries are subsequently transferred to the ionization continuum through further photon absorption, although generally with a different phase dependence from that of the excited-state populations, highlighting the influence of multiple competing pathways on the final ion yields.

The observation of large asymmetries in the parent ions is mirrored by significant dichroism in the fragment ions. Interestingly, for several fragments, the enantio-sensitive response peaks at phases distinct from those of the parent ions, suggesting that fragmentation may be driven by a different ionization channel. More broadly, these results illustrate the potential of synthetic chiral light as a sensitive probe of quantum pathways, electronic currents, and chiral dynamics,\cite{smirnova2025} while also offering new opportunities to induce and control chiral and spin-dependent phenomena,\cite{bloom2024} and to drive selective chiral photochemistry\cite{moore2024}.

Ultimately, the ability to synthesize chiral light from multiple laser fields with independently tunable properties opens a vast multi-dimensional control space. Exploring and optimizing this space, including through machine-learning approaches, may enable substantially enhanced chiral discrimination and unprecedented control over light-matter interactions in complex molecular systems. Unlike conventional asymmetric photochemistry, which relies on chiral catalysts to induce enantioselectivity, synthetic chiral light achieves this optically - opening the possibility of enantioselective control at optical spatial and temporal scales, with potential applications for chiroptical molecular switches and motors \cite{huck1996,feringa2007,wachsmuth2025}.

%
%
%


\clearpage 

%
\bibliography{biblio} 

@article{cormier1996,
doi = {10.1088/0953-4075/29/9/013},
year = {1996},
month = {may},
publisher = {},
volume = {29},
number = {9},
pages = {1667},
author = {E Cormier and P Lambropoulos},
title = {Optimal gauge and gauge invariance in non-perturbative time-dependent calculation of above-threshold ionization},
journal = {Journal of Physics B: Atomic, Molecular and Optical Physics}
}

@misc{pubchem82229,
  author       = {{National Center for Biotechnology Information}},
  title        = {{PubChem Compound Summary for CID 82229, (-)-Fenchone}},
  howpublished = {\url{https://pubchem.ncbi.nlm.nih.gov/compound/L-Fenchone}}
}

@Article{kohndensityfunctionaltheory1996,
  author	= {Kohn, W. and Becke, A. D. and Parr, R. G.},
  journal	= {The Journal of Physical Chemistry},
  month		= jan,
  number	= {31},
  pages		= {12974--12980},
  publisher	= {American Chemical Society},
  title		= {Density {{Functional Theory}} of {{Electronic
		  Structure}}},
  volume	= {100},
  year		= {1996},
  doi		= {10.1021/jp960669l},
  issn		= {0022-3654}
}

@Article{	  leedevelopmentcollesalvetticorrelationenergy1988,
  author	= {Lee, Chengteh and Yang, Weitao and Parr, Robert G.},
  journal	= {Physical Review B},
  month		= jan,
  number	= {2},
  pages		= {785--789},
  publisher	= {American Physical Society},
  title		= {Development of the {{Colle-Salvetti}} Correlation-Energy
		  Formula into a Functional of the Electron Density},
  volume	= {37},
  year		= {1988},
  doi		= {10.1103/PhysRevB.37.785}
}

@Article{beckedensityfunctionalexchangeenergyapproximation1988,
  author	= {Becke, A. D.},
  journal	= {Physical Review A},
  month		= sep,
  number	= {6},
  pages		= {3098--3100},
  publisher	= {American Physical Society},
  title		= {Density-Functional Exchange-Energy Approximation with
		  Correct Asymptotic Behavior},
  volume	= {38},
  year		= {1988},
  doi		= {10.1103/PhysRevA.38.3098}
}

@Article{	 mcleancontractedgaussianbasis2008,
  author	= {McLean, A. D. and Chandler, G. S.},
  journal	= {The Journal of Chemical Physics},
  month		= jul,
  number	= {10},
  pages		= {5639--5648},
  title		= {Contracted {{Gaussian}} Basis Sets for Molecular
		  Calculations. {{I}}. {{Second}} Row Atoms, {{Z}}=11--18},
  volume	= {72},
  year		= {2008},
  doi		= {10.1063/1.438980},
  issn		= {0021-9606}
}

@article{Krishnan80,
    author = {Krishnan, R. and Binkley, J. S. and Seeger, R. and Pople, J. A.},
    title = {Self‐consistent molecular orbital methods. XX. A basis set for correlated wave functions},
    journal = {The Journal of Chemical Physics},
    volume = {72},
    number = {1},
    pages = {650-654},
    year = {1980},
    issn = {0021-9606},
    doi = {10.1063/1.438955},
}

@article{ORCA6,
author = {Neese, Frank},
title = {Software Update: The ORCA Program System—Version 6.0},
journal = {Wiley Interdiscip. Rev. Comput. Mol. Sci.},
volume = {15},
number = {2},
pages = {e70019},
doi = {10.1002/wcms.70019},
year = {2025}
}

@inbook{TDDFT,
author = {MARK E. CASIDA},
title = {Time-Dependent Density Functional Response Theory for Molecules},
booktitle = {Recent Advances in Density Functional Methods},
chapter = {},
pages = {155-192},
doi = {10.1142/9789812830586_0005},
eprint = {https://www.worldscientific.com/doi/pdf/10.1142/9789812830586_0005}
}

@article{YANAI200451,
title = {A new hybrid exchange–correlation functional using the Coulomb-attenuating method (CAM-B3LYP)},
journal = {Chemical Physics Letters},
volume = {393},
number = {1},
pages = {51-57},
year = {2004},
issn = {0009-2614},
doi = {https://doi.org/10.1016/j.cplett.2004.06.011},
author = {Takeshi Yanai and David P Tew and Nicholas C Handy}
}

@article{dunning89,
    author = {Dunning, Thom H., Jr.},
    title = {Gaussian basis sets for use in correlated molecular calculations. I. The atoms boron through neon and hydrogen},
    journal = {The Journal of Chemical Physics},
    volume = {90},
    number = {2},
    pages = {1007-1023},
    year = {1989},
    month = {01},
    issn = {0021-9606},
    doi = {10.1063/1.456153},
    eprint = {https://pubs.aip.org/aip/jcp/article-pdf/90/2/1007/18974738/1007_1_online.pdf},
}

@article{augccpvdz,
    author = {Papajak, Ewa and Zheng, Jingjing and Xu, Xuefei and Leverentz, Hannah R. and Truhlar, Donald G.},
    title = {Perspectives on Basis Sets Beautiful: Seasonal Plantings of Diffuse Basis Functions},
    journal = {Journal of Chemical Theory and Computation},
    volume = {7},
    number = {10},
    pages = {3027-3034},
    year = {2011},
    month = {08},
    issn = {1549-9618},
    doi = {10.1021/ct200106a},
    eprint = {https://pubs.acs.org/jctcce/article-pdf/7/10/3027/16650770/ct200106a.pdf},
}

@article{QChem,
author = {Yihan Shao and Zhengting Gan and Evgeny Epifanovsky and Andrew T.B. Gilbert and Michael Wormit and Joerg Kussmann and Adrian W. Lange and Andrew Behn and Jia Deng and Xintian Feng and Debashree Ghosh and Matthew Goldey and Paul R. Horn and Leif D. Jacobson and Ilya Kaliman and Rustam Z. Khaliullin and Tomasz Kuś and Arie Landau and Jie Liu and Emil I. Proynov and Young Min Rhee and Ryan M. Richard and Mary A. Rohrdanz and Ryan P. Steele and Eric J. Sundstrom and H. Lee Woodcock III and Paul M. Zimmerman and Dmitry Zuev and Ben Albrecht and Ethan Alguire and Brian Austin and Gregory J. O. Beran and Yves A. Bernard and Eric Berquist and Kai Brandhorst and Ksenia B. Bravaya and Shawn T. Brown and David Casanova and Chun-Min Chang and Yunqing Chen and Siu Hung Chien and Kristina D. Closser and Deborah L. Crittenden and Michael Diedenhofen and Robert A. DiStasio Jr. and Hainam Do and Anthony D. Dutoi and Richard G. Edgar and Shervin Fatehi and Laszlo Fusti-Molnar and An Ghysels and Anna Golubeva-Zadorozhnaya and Joseph Gomes and Magnus W.D. Hanson-Heine and Philipp H.P. Harbach and Andreas W. Hauser and Edward G. Hohenstein and Zachary C. Holden and Thomas-C. Jagau and Hyunjun Ji and Benjamin Kaduk and Kirill Khistyaev and Jaehoon Kim and Jihan Kim and Rollin A. King and Phil Klunzinger and Dmytro Kosenkov and Tim Kowalczyk and Caroline M. Krauter and Ka Un Lao and Adèle D. Laurent and Keith V. Lawler and Sergey V. Levchenko and Ching Yeh Lin and Fenglai Liu and Ester Livshits and Rohini C. Lochan and Arne Luenser and Prashant Manohar and Samuel F. Manzer and Shan-Ping Mao and Narbe Mardirossian and Aleksandr V. Marenich and Simon A. Maurer and Nicholas J. Mayhall and Eric Neuscamman and C. Melania Oana and Roberto Olivares-Amaya and Darragh P. O’Neill and John A. Parkhill and Trilisa M. Perrine and Roberto Peverati and Alexander Prociuk and Dirk R. Rehn and Edina Rosta and Nicholas J. Russ and Shaama M. Sharada and Sandeep Sharma and David W. Small and Alexander Sodt and Tamar Stein and David Stück and Yu-Chuan Su and Alex J.W. Thom and Takashi Tsuchimochi and Vitalii Vanovschi and Leslie Vogt and Oleg Vydrov and Tao Wang and Mark A. Watson and Jan Wenzel and Alec White and Christopher F. Williams and Jun Yang and Sina Yeganeh and Shane R. Yost and Zhi-Qiang You and Igor Ying Zhang and Xing Zhang and Yan Zhao and Bernard R. Brooks and Garnet K.L. Chan and Daniel M. Chipman and Christopher J. Cramer and William A. Goddard III and Mark S. Gordon and Warren J. Hehre and Andreas Klamt and Henry F. Schaefer III and Michael W. Schmidt and C. David Sherrill and Donald G. Truhlar and Arieh Warshel and Xin Xu and Alán Aspuru-Guzik and Roi Baer and Alexis T. Bell and Nicholas A. Besley and Jeng-Da Chai and Andreas Dreuw and Barry D. Dunietz and Thomas R. Furlani and Steven R. Gwaltney and Chao-Ping Hsu and Yousung Jung and Jing Kong and Daniel S. Lambrecht and WanZhen Liang and Christian Ochsenfeld and Vitaly A. Rassolov and Lyudmila V. Slipchenko and Joseph E. Subotnik and Troy Van Voorhis and John M. Herbert and Anna I. Krylov and Peter M.W. Gill and Martin Head-Gordon},
title = {Advances in molecular quantum chemistry contained in the Q-Chem 4 program package},
journal = {Molecular Physics},
volume = {113},
number = {2},
pages = {184--215},
year = {2015},
publisher = {Taylor \& Francis},
doi = {10.1080/00268976.2014.952696},
 }

@article{ayuso2019,
  title = {Synthetic Chiral Light for Efficient Control of Chiral Light--Matter Interaction},
  author = {Ayuso, David and Neufeld, Ofer and Ordonez, Andres F. and Decleva, Piero and Lerner, Gavriel and Cohen, Oren and Ivanov, Misha and Smirnova, Olga},
  year = 2019,
  month = oct,
  journal = {Nature Photonics},
  pages = {1--6},
  issn = {1749-4893},
  doi = {10.1038/s41566-019-0531-2},
  urldate = {2019-11-05},
  copyright = {2019 The Author(s), under exclusive licence to Springer Nature Limited},
  langid = {english}
}

@article{ayuso2022,
  title = {Ultrafast Chirality: The Road to Efficient Chiral Measurements},
  shorttitle = {Ultrafast Chirality},
  author = {Ayuso, David and Ordonez, Andres F. and Smirnova, Olga},
  year = 2022,
  month = nov,
  journal = {Physical Chemistry Chemical Physics},
  volume = {24},
  number = {44},
  pages = {26962--26991},
  publisher = {The Royal Society of Chemistry},
  issn = {1463-9084},
  doi = {10.1039/D2CP01009G},
  urldate = {2023-05-12},
  langid = {english}
}

@article{beaulieu2024,
  title = {Strong-Field Ionization of Chiral Molecules with Bicircular Laser Fields: {{Sub-barrier}} Dynamics, Interference, and Vortices},
  shorttitle = {Strong-Field Ionization of Chiral Molecules with Bicircular Laser Fields},
  author = {Beaulieu, S. and Larroque, S. and Descamps, D. and Fabre, B. and Petit, S. and Ta{\"i}eb, R. and Pons, B. and Mairesse, Y.},
  year = 2024,
  month = jul,
  journal = {Physical Review A},
  volume = {110},
  number = {1},
  pages = {013103},
  publisher = {American Physical Society},
  doi = {10.1103/PhysRevA.110.013103},
  urldate = {2024-07-02}
}

@book{berova2000,
  title = {Circular Dichroism Principles and Applications},
  author = {Berova, N. and Nakanishi, R.},
  year = 2000,
  publisher = {Wiley-VCH, New York}
}

@article{blanchet2021,
  title = {Ultrafast Relaxation Investigated by Photoelectron Circular Dichroism: An Isomeric Comparison of Camphor and Fenchone},
  shorttitle = {Ultrafast Relaxation Investigated by Photoelectron Circular Dichroism},
  author = {Blanchet, Val{\'e}rie and Descamps, Dominique and Petit, St{\'e}phane and Mairesse, Yann and Pons, Bernard and Fabre, Baptiste},
  year = 2021,
  month = nov,
  journal = {Physical Chemistry Chemical Physics},
  volume = {23},
  number = {45},
  pages = {25612--25628},
  publisher = {The Royal Society of Chemistry},
  issn = {1463-9084},
  doi = {10.1039/D1CP03569J},
  urldate = {2021-11-25},
  langid = {english}
}

@article{bloch2021,
  title = {Revealing the {{Influence}} of {{Molecular Chirality}} on {{Tunnel-Ionization Dynamics}}},
  author = {Bloch, E. and Larroque, S. and Rozen, S. and Beaulieu, S. and Comby, A. and Beauvarlet, S. and Descamps, D. and Fabre, B. and Petit, S. and Ta{\"i}eb, R. and Uzan, A. J. and Blanchet, V. and Dudovich, N. and Pons, B. and Mairesse, Y.},
  year = 2021,
  month = dec,
  journal = {Physical Review X},
  volume = {11},
  number = {4},
  pages = {041056},
  publisher = {American Physical Society},
  doi = {10.1103/PhysRevX.11.041056},
  urldate = {2021-12-23}
}

@article{bloom2024,
  title = {Chiral {{Induced Spin Selectivity}}},
  author = {Bloom, Brian P. and Paltiel, Yossi and Naaman, Ron and Waldeck, David H.},
  year = 2024,
  month = feb,
  journal = {Chemical Reviews},
  volume = {124},
  number = {4},
  pages = {1950--1991},
  issn = {0009-2665},
  doi = {10.1021/acs.chemrev.3c00661},
  urldate = {2026-08-12}
}

@article{bowering2001,
  title = {Asymmetry in {{Photoelectron Emission}} from {{Chiral Molecules Induced}} by {{Circularly Polarized Light}}},
  author = {B{\"o}wering, N. and Lischke, T. and Schmidtke, B. and M{\"u}ller, N. and Khalil, T. and Heinzmann, U.},
  year = 2001,
  month = feb,
  journal = {Physical Review Letters},
  volume = {86},
  number = {7},
  pages = {1187--1190},
  doi = {10.1103/PhysRevLett.86.1187},
  urldate = {2014-02-27}
}

@article{calegari2014,
  title = {Ultrafast Electron Dynamics in Phenylalanine Initiated by Attosecond Pulses},
  author = {Calegari, F. and Ayuso, D. and Trabattoni, A. and Belshaw, L. and Camillis, S. De and Anumula, S. and Frassetto, F. and Poletto, L. and Palacios, A. and Decleva, P. and Greenwood, J. B. and Mart{\'i}n, F. and Nisoli, M.},
  year = 2014,
  month = oct,
  journal = {Science},
  volume = {346},
  number = {6207},
  pages = {336--339},
  issn = {0036-8075, 1095-9203},
  doi = {10.1126/science.1254061},
  urldate = {2015-02-23},
  langid = {english},
  pmid = {25324385}
}

@article{cameron2023,
  title = {Strong Chiral Optical Force for Small Chiral Molecules Based on Electric-Dipole Interactions, Inspired by the Asymmetrical Hydrozoan {{Velella}} Velella},
  author = {Cameron, Robert P and McArthur, Duncan and Yao, Alison M},
  year = 2023,
  month = aug,
  journal = {New Journal of Physics},
  volume = {25},
  number = {8},
  pages = {083006},
  publisher = {IOP Publishing},
  issn = {1367-2630},
  doi = {10.1088/1367-2630/ace7ee},
  urldate = {2026-04-23},
  langid = {english}
}

@article{comby2016,
  title = {Relaxation {{Dynamics}} in {{Photoexcited Chiral Molecules Studied}} by {{Time-Resolved Photoelectron Circular Dichroism}}: {{Toward Chiral Femtochemistry}}},
  shorttitle = {Relaxation {{Dynamics}} in {{Photoexcited Chiral Molecules Studied}} by {{Time-Resolved Photoelectron Circular Dichroism}}},
  author = {Comby, Antoine and Beaulieu, Samuel and {Boggio-Pasqua}, Martial and Descamps, Dominique and L{\'e}gar{\'e}, Francois and Nahon, Laurent and Petit, St{\'e}phane and Pons, Bernard and Fabre, Baptiste and Mairesse, Yann and Blanchet, Val{\'e}rie},
  year = 2016,
  month = nov,
  journal = {The Journal of Physical Chemistry Letters},
  volume = {7},
  number = {22},
  pages = {4514--4519},
  issn = {1948-7185},
  doi = {10.1021/acs.jpclett.6b02065},
  urldate = {2017-02-02}
}

@article{comby2018a,
  title = {Real-Time Determination of Enantiomeric and Isomeric Content Using Photoelectron Elliptical Dichroism},
  author = {Comby, A. and Bloch, E. and Bond, C. M. M. and Descamps, D. and Miles, J. and Petit, S. and Rozen, S. and Greenwood, J. B. and Blanchet, V. and Mairesse, Y.},
  year = 2018,
  month = dec,
  journal = {Nature Communications},
  volume = {9},
  number = {1},
  pages = {5212},
  issn = {2041-1723},
  doi = {10.1038/s41467-018-07609-9},
  urldate = {2018-12-06},
  copyright = {2018 The Author(s)},
  langid = {english}
}

@article{eibenberger2017,
  title = {Enantiomer-{{Specific State Transfer}} of {{Chiral Molecules}}},
  author = {Eibenberger, Sandra and Doyle, John and Patterson, David},
  year = 2017,
  month = mar,
  journal = {Physical Review Letters},
  volume = {118},
  number = {12},
  pages = {123002},
  doi = {10.1103/PhysRevLett.118.123002},
  urldate = {2017-03-27}
}

@article{feringa2007,
  title = {The {{Art}} of {{Building Small}}:\, {{From Molecular Switches}} to {{Molecular Motors}}},
  shorttitle = {The {{Art}} of {{Building Small}}},
  author = {Feringa, Ben L.},
  year = 2007,
  month = jul,
  journal = {The Journal of Organic Chemistry},
  volume = {72},
  number = {18},
  pages = {6635--6652},
  issn = {0022-3263},
  doi = {10.1021/jo070394d},
  urldate = {2026-09-01}
}

@article{frishman2003,
  title = {Enantiomeric Purification of Nonpolarized Racemic Mixtures Using Coherent Light},
  author = {Frishman, Einat and Shapiro, Moshe and Gerbasi, David and Brumer, Paul},
  year = 2003,
  month = oct,
  journal = {The Journal of Chemical Physics},
  volume = {119},
  number = {14},
  pages = {7237--7246},
  issn = {0021-9606},
  doi = {10.1063/1.1603732},
  urldate = {2026-08-10}
}

@article{greenwood2023,
  title = {Investigation of Photoelectron Elliptical Dichroism for Chiral Analysis},
  author = {Greenwood, Jason B. and Williams, Ian D.},
  year = 2023,
  month = jun,
  journal = {Physical Chemistry Chemical Physics},
  volume = {25},
  number = {24},
  pages = {16238--16245},
  publisher = {The Royal Society of Chemistry},
  issn = {1463-9084},
  doi = {10.1039/D3CP01058A},
  urldate = {2023-06-21},
  langid = {english}
}

@article{greenwood2025,
  title = {Quantum Control of Photoion Circular Dichroism Using Orthogonal Laser Beams},
  author = {Greenwood, Jason B. and Donnelly, Leah},
  year = 2025,
  month = may,
  journal = {Physical Chemistry Chemical Physics},
  publisher = {The Royal Society of Chemistry},
  issn = {1463-9084},
  doi = {10.1039/D5CP01127B},
  urldate = {2025-05-22},
  langid = {english}
}

@article{habibovic2024,
  title = {Emerging Tailored Light Sources for Studying Chirality and Symmetry},
  author = {Habibovi{\'c}, Dino and Hamilton, Kathryn R. and Neufeld, Ofer and Rego, Laura},
  year = 2024,
  month = nov,
  journal = {Nature Reviews Physics},
  volume = {6},
  number = {11},
  pages = {663--675},
  publisher = {Nature Publishing Group},
  issn = {2522-5820},
  doi = {10.1038/s42254-024-00763-8},
  urldate = {2026-06-16},
  copyright = {2024 Springer Nature Limited},
  langid = {english}
}

@article{huck1996,
  title = {Dynamic {{Control}} and {{Amplification}} of {{Molecular Chirality}} by {{Circular Polarized Light}}},
  author = {Huck, Nina P. M. and Jager, Wolter F. and {de Lange}, Ben and Feringa, Ben L.},
  year = 1996,
  month = sep,
  journal = {Science},
  volume = {273},
  number = {5282},
  pages = {1686--1688},
  publisher = {American Association for the Advancement of Science},
  doi = {10.1126/science.273.5282.1686},
  urldate = {2026-09-01}
}

@article{kohnke2026,
  title = {Multiphoton Ionization with Three-Dimensional Light Fields},
  author = {K{\"o}hnke, D. and Ahlswede, H.-C. and Bayer, T. and Wollenhaupt, M.},
  year = 2026,
  month = jul,
  journal = {Physical Review Research},
  volume = {8},
  number = {3},
  pages = {033048},
  publisher = {American Physical Society},
  doi = {10.1103/r36b-vw82},
  urldate = {2026-09-01}
}

@article{lee2022,
  title = {Quantitative {{Study}} of {{Enantiomer-Specific State Transfer}}},
  author = {Lee, JuHyeon and Bischoff, Johannes and {Hernandez-Castillo}, A. O. and Sartakov, Boris and Meijer, Gerard and {Eibenberger-Arias}, Sandra},
  year = 2022,
  month = apr,
  journal = {Physical Review Letters},
  volume = {128},
  number = {17},
  pages = {173001},
  publisher = {American Physical Society},
  doi = {10.1103/PhysRevLett.128.173001},
  urldate = {2022-10-04}
}

@article{leibscher2019,
  title = {Principles of Enantio-Selective Excitation in Three-Wave Mixing Spectroscopy of Chiral Molecules},
  author = {Leibscher, Monika and Giesen, Thomas F. and Koch, Christiane P.},
  year = 2019,
  month = jul,
  journal = {The Journal of Chemical Physics},
  volume = {151},
  number = {1},
  pages = {014302},
  issn = {0021-9606},
  doi = {10.1063/1.5097406},
  urldate = {2026-08-10}
}

@article{liu2023a,
  title = {Detection and Analysis of Chiral Molecules as Disease Biomarkers},
  author = {Liu, Yaoran and Wu, Zilong and Armstrong, Daniel W. and Wolosker, Herman and Zheng, Yuebing},
  year = 2023,
  month = may,
  journal = {Nature Reviews Chemistry},
  volume = {7},
  number = {5},
  pages = {355--373},
  publisher = {Nature Publishing Group},
  issn = {2397-3358},
  doi = {10.1038/s41570-023-00476-z},
  urldate = {2026-06-16},
  copyright = {2023 Springer Nature Limited},
  langid = {english}
}

@article{miles2017,
  title = {A New Technique for Probing Chirality via Photoelectron Circular Dichroism},
  author = {Miles, J. and Fernandes, D. and Young, A. and Bond, C. M. M. and Crane, S. W. and Ghafur, O. and Townsend, D. and S{\'a}, J. and Greenwood, J. B.},
  year = 2017,
  month = sep,
  journal = {Analytica Chimica Acta},
  volume = {984},
  pages = {134--139},
  issn = {0003-2670},
  doi = {10.1016/j.aca.2017.06.051},
  urldate = {2018-07-06}
}

@article{moore2024,
  title = {Significant {{Chiral Asymmetry Observed}} in {{Neutral Amino Acid Ultraviolet Photolysis}}},
  author = {Moore, Brendan and Zeng, Linshan and Djuricanin, Pavle and Cooke, Ilsa R. and Madison, Kirk W. and Momose, Takamasa},
  year = 2024,
  month = dec,
  journal = {Journal of the American Chemical Society},
  volume = {146},
  number = {50},
  pages = {34333--34340},
  publisher = {American Chemical Society},
  issn = {0002-7863},
  doi = {10.1021/jacs.4c07585},
  urldate = {2026-06-16}
}

@article{nahon2015,
  title = {Valence Shell One-Photon Photoelectron Circular Dichroism in Chiral Systems},
  author = {Nahon, Laurent and Garcia, Gustavo A. and Powis, Ivan},
  year = 2015,
  month = oct,
  journal = {Journal of Electron Spectroscopy and Related Phenomena},
  series = {Gas Phase Spectroscopic and Dynamical Studies at {{Free-Electron Lasers}} and Other Short Wavelength Sources},
  volume = {204, Part B},
  pages = {322--334},
  issn = {0368-2048},
  doi = {10.1016/j.elspec.2015.04.008},
  urldate = {2016-05-02}
}

@article{neufeld2021,
  title = {Strong Chiral Dichroism and Enantiopurification in Above-Threshold Ionization with Locally Chiral Light},
  author = {Neufeld, Ofer and H{\"u}bener, Hannes and Rubio, Angel and De Giovannini, Umberto},
  year = 2021,
  month = jul,
  journal = {Physical Review Research},
  volume = {3},
  number = {3},
  pages = {L032006},
  publisher = {American Physical Society},
  doi = {10.1103/PhysRevResearch.3.L032006},
  urldate = {2021-07-08}
}

@article{ordonez2018,
  title = {Generalized Perspective on Chiral Measurements without Magnetic Interactions},
  author = {Ordonez, Andres F. and Smirnova, Olga},
  year = 2018,
  month = dec,
  journal = {Physical Review A},
  volume = {98},
  number = {6},
  pages = {063428},
  doi = {10.1103/PhysRevA.98.063428},
  urldate = {2019-01-22}
}

@article{ordonez2025a,
  title = {Shaping Light in {{3D}} for Direct Asymmetric Photochemistry: A 1000-Fold Enantioselectivity Enhancement},
  shorttitle = {Shaping Light in {{3D}} for Direct Asymmetric Photochemistry},
  author = {Ord{\'o}{\~n}ez, Andr{\'e}s and Garc{\'i}a Gonz{\'a}lez, Daniel and {Vindel-Zandbergen}, Patricia and Ayuso, David},
  year = 2026,
  month = sep,
  journal = {Nature Communications},
  publisher = {Nature Publishing Group},
  issn = {2041-1723},
  doi = {10.1038/s41467-026-77096-w},
  urldate = {2026-09-04},
  copyright = {2026 The Author(s)},
  langid = {english},
}

@article{perez2017,
  title = {Coherent {{Enantiomer-Selective Population Enrichment Using Tailored Microwave Fields}}},
  author = {P{\'e}rez, Crist{\'o}bal and Steber, Amanda L. and Domingos, S{\'e}rgio R. and Krin, Anna and Schmitz, David and Schnell, Melanie},
  year = 2017,
  journal = {Angewandte Chemie International Edition},
  volume = {56},
  number = {41},
  pages = {12512--12517},
  issn = {1521-3773},
  doi = {10.1002/anie.201704901},
  urldate = {2026-08-10},
  copyright = {\copyright{} 2017 Wiley-VCH Verlag GmbH \& Co. KGaA, Weinheim},
  langid = {english}
}

@article{pitzer2013,
  title = {Direct {{Determination}} of {{Absolute Molecular Stereochemistry}} in {{Gas Phase}} by {{Coulomb Explosion Imaging}}},
  author = {Pitzer, Martin and Kunitski, Maksim and Johnson, Allan S. and Jahnke, Till and Sann, Hendrik and Sturm, Felix and Schmidt, Lothar Ph H. and {Schmidt-B{\"o}cking}, Horst and D{\"o}rner, Reinhard and Stohner, J{\"u}rgen and Kiedrowski, Julia and Reggelin, Michael and Marquardt, Sebastian and Schie{\ss}er, Alexander and Berger, Robert and Sch{\"o}ffler, Markus S.},
  year = 2013,
  month = jun,
  journal = {Science},
  volume = {341},
  number = {6150},
  pages = {1096--1100},
  issn = {0036-8075, 1095-9203},
  doi = {10.1126/science.1240362},
  urldate = {2013-09-17},
  langid = {english},
  pmid = {24009390}
}

@article{rego2023,
  title = {Structuring the Local Handedness of Synthetic Chiral Light: Global Chirality versus Polarization of Chirality},
  shorttitle = {Structuring the Local Handedness of Synthetic Chiral Light},
  author = {Rego, Laura and Ayuso, David},
  year = 2023,
  month = sep,
  journal = {New Journal of Physics},
  volume = {25},
  number = {9},
  pages = {093005},
  publisher = {IOP Publishing},
  issn = {1367-2630},
  doi = {10.1088/1367-2630/acf150},
  urldate = {2026-06-16},
  langid = {english}
}

@article{ritchie1975,
  title = {Theoretical Studies in Photoelectron Spectroscopy. {{Molecular}} Optical Activity in the Region of Continuous Absorption and Its Characterization by the Angular Distribution of Photoelectrons},
  author = {Ritchie, Burke},
  year = 1975,
  month = aug,
  journal = {Physical Review A},
  volume = {12},
  number = {2},
  pages = {567--574},
  publisher = {American Physical Society},
  doi = {10.1103/PhysRevA.12.567},
  urldate = {2022-06-29}
}

@article{romao2025,
  title = {Chirality \`a La Carte},
  author = {Romao, Carl P. and Juraschek, Dominik M.},
  year = 2025,
  month = jan,
  journal = {Science},
  volume = {387},
  number = {6732},
  pages = {361--362},
  publisher = {American Association for the Advancement of Science},
  doi = {10.1126/science.adv0319},
  urldate = {2026-06-16}
}

@article{rozen2019,
  title = {Controlling {{Subcycle Optical Chirality}} in the {{Photoionization}} of {{Chiral Molecules}}},
  author = {Rozen, S. and Comby, A. and Bloch, E. and Beauvarlet, S. and Descamps, D. and Fabre, B. and Petit, S. and Blanchet, V. and Pons, B. and Dudovich, N. and Mairesse, Y.},
  year = 2019,
  month = jul,
  journal = {Physical Review X},
  volume = {9},
  number = {3},
  pages = {031004},
  doi = {10.1103/PhysRevX.9.031004},
  urldate = {2019-07-08}
}

@article{smirnova2025,
  title = {A New Age of Molecular Chirality},
  author = {Smirnova, Olga},
  year = 2025,
  month = jul,
  journal = {Science},
  volume = {389},
  number = {6757},
  pages = {232--233},
  publisher = {American Association for the Advancement of Science},
  doi = {10.1126/science.adn0905},
  urldate = {2025-11-06}
}

@article{sparling2025,
  title = {Two Decades of Imaging Photoelectron Circular Dichroism: From First Principles to Future Perspectives},
  shorttitle = {Two Decades of Imaging Photoelectron Circular Dichroism},
  author = {Sparling, Chris and Townsend, Dave},
  year = 2025,
  month = feb,
  journal = {Physical Chemistry Chemical Physics},
  volume = {27},
  number = {6},
  pages = {2888--2907},
  issn = {1463-9076},
  doi = {10.1039/d4cp03770g},
  urldate = {2026-08-12}
}

@article{wachsmuth2025,
  title = {A Molecular Machine Directs the Synthesis of a Catenane},
  author = {Wachsmuth, Tommy and Kluifhooft, Robert and M{\"u}ller, Mira and Zei{\ss}, Leon and Kathan, Michael},
  year = 2025,
  month = jul,
  journal = {Science},
  volume = {389},
  number = {6759},
  pages = {526--531},
  publisher = {American Association for the Advancement of Science},
  doi = {10.1126/science.adx5363},
  urldate = {2026-09-03}
}

@article{ye2023,
  title = {Phase-Matched Locally Chiral Light for Global Control of Chiral Light--Matter Interaction},
  author = {Ye, Chong and Sun, Yifan and Fu, Libin and Zhang, Xiangdong},
  year = 2023,
  month = nov,
  journal = {Optics Letters},
  volume = {48},
  number = {21},
  pages = {5511--5514},
  publisher = {Optica Publishing Group},
  issn = {1539-4794},
  doi = {10.1364/OL.496226},
  urldate = {2026-06-16},
  copyright = {\copyright{} 2023 Optica Publishing Group},
  langid = {english}
}

@article{zeng2025,
  title = {Photo-Induced Chirality in a Nonchiral Crystal},
  author = {Zeng, Z. and F{\"o}rst, M. and Fechner, M. and Buzzi, M. and Amuah, E. B. and Putzke, C. and Moll, P. J. W. and Prabhakaran, D. and Radaelli, P. G. and Cavalleri, A.},
  year = 2025,
  month = jan,
  journal = {Science},
  volume = {387},
  number = {6732},
  pages = {431--436},
  publisher = {American Association for the Advancement of Science},
  doi = {10.1126/science.adr4713},
  urldate = {2025-01-29}
}
\bibliographystyle{sciencemag}

%
%
%
%
%
%


\section*{Acknowledgments}
We thank Franck Blais, Rodrigue Bouillaud, Romain Delos, Nikita Fedorov and Laurent Merzeau for technical assistance. We acknowledge Samuel Beaulieu and Baptiste Fabre  for fruitful discussions. 
\paragraph*{Funding:}
J.G. acknowledges support from the Royal Society’s Paul Instrument Fund (PI170043) and International Exchange Scheme (IES \textbackslash R1\textbackslash 241228), and from Queen’s University’s Agility Plus fund. A.O. and D.G.G. acknowledge funding from the Deutsche Forschungsgemeinschaft (DFG, German Research Foundation) 543760364. L.F. acknowledges the financial support of the IdEx University of Bordeaux / Grand Research ``GPR LIGHT". D.A. acknowledges funding from the Royal Society URF\textbackslash R\textbackslash251036.
R.A. and C.S. thank Heriot-Watt University for PhD funding through provision of an EPSRC Doctoral Training Partnership award, as well as travel support through the Outbound Research Programme. This work is part of the ULTRAFAST and TORNADO projects of PEPR LUMA and was supported by the French National Research Agency, under grants ANR-21-CE30-038-01 (Shotime) and ANR-24-CE92-0039 (CACTUS), and as a part of the France 2030 program, under grants ANR-23-EXLU-0002 and ANR-23-EXLU-0004. This project has received funding from European Union's Horizon  research and innovation programme under ERC Starting Grant ERC-2022-STG No.101076639, under the Marie Skłodowska-Curie grants No HORIZON-MSCA-2023-DN-01 101169225 - SPARKLE and HORIZON-MSCA-2024-DN-01 101226887 - TRILOGY, and under HE-GA 101131771 Lasers4EU [PID:41758].  Views and opinions expressed are however those of the authors only and do not necessarily reflect those of the European Union. Neither the European Union nor the granting authority can be held responsible for them. 
\paragraph*{Author contributions:}
J.G. conceived the study, with A.O. and D.A. providing theoretical insight. J.G., L.F., U.D., V.B. and Y.M. built the experiment. J.G., R.A. and Y.M. carried out the measurements. J.G. and Y.M. analyzed the results. A.O., D.G.G. and B.P. carried out the simulations. P.V.Z. and D.A. calculated energies and transition dipoles for fenchone. J.G. and Y.M. wrote the article with contributions from A.O. and B.P. All authors provided input to the content and conclusions.

\paragraph*{Competing interests:}
There are no competing interests to declare.

\paragraph*{Data, code and materials availability:}
\subsection*{Supplementary materials}
Materials and Methods\\
Supplementary Text\\
Figs. S1 to S12\\
Tables S1 to S5\\
References \textit{(34-\arabic{enumiv})}\\ 


\newpage


\renewcommand{\thefigure}{S\arabic{figure}}
\renewcommand{\thetable}{S\arabic{table}}
\renewcommand{\theequation}{S\arabic{equation}}
\renewcommand{\thepage}{S\arabic{page}}
\setcounter{figure}{0}
\setcounter{table}{0}
\setcounter{equation}{0}
\setcounter{page}{1} 


\begin{center}
\section*{Supplementary Materials for\\ \scititle}


	Jason B. Greenwood$^{\ast}$, Letizia Fede, Daniel García González, \and 
	Umberto Dellasette, Ross Anderson, Chris Sparling, Dave Townsend, \and
    Patricia Vindel-Zandbergen, David Ayuso, Val\'erie Blanchet, \and 
    Bernard Pons, Andr\'es Ord\'o\~nez,
	Yann Mairesse\and

	\small$^\ast$Corresponding author. Email:  j.greenwood@qub.ac.uk\and

\end{center}

\subsubsection*{This PDF file includes:}
Materials and Methods\\
Figures S1 to S12\\
Tables S1 to S5\\

\newpage


\subsection*{Materials and Methods}




\subsubsection*{1. EXPERIMENTAL METHODS}

\subsubsection*{1.1 Optical Arrangement}
The laser source used for the measurements was a Tangor (Amplitude) delivering 450~fs pulses at 1030 nm central wavelength, at a repetition rate of 200~kHz. The laser pulses were post-compressed in a multipass cell filled with 2.7~bar of xenon, followed by a set of chirped mirrors, resulting in 80 fs pulses with 150~$\mu$J energy. 

\begin{figure}
    \centering
    \includegraphics[width=1\linewidth]{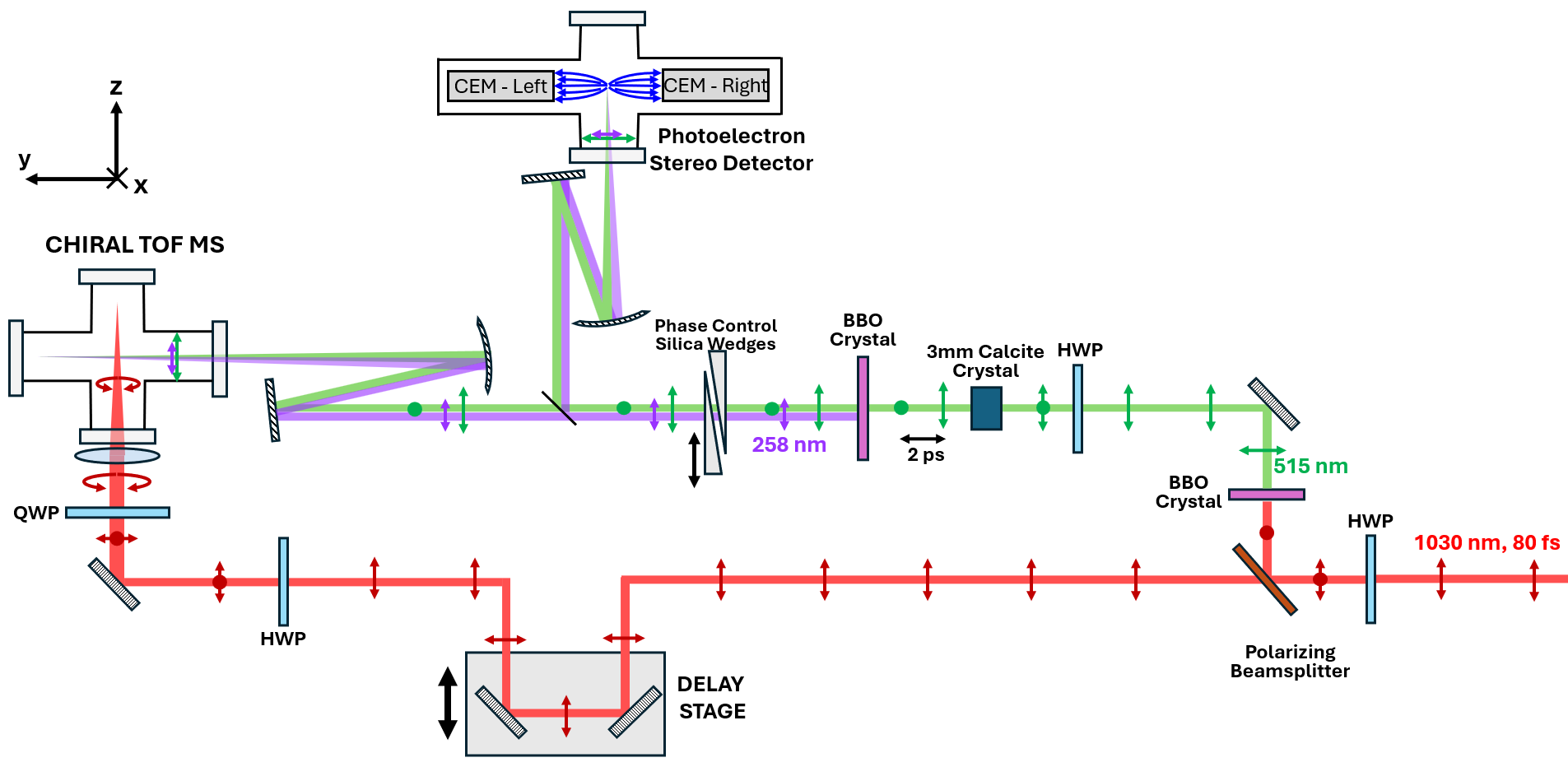}
    \caption{\textbf{Optics used to synthesize chiral light in the $E_{x\pm iy,z,z}$ configuration.} The colors red, green and lilac show the 
    1$\omega$, 2$\omega$ and 4$\omega$ beams respectively, with arrows and dots showing horizontal (\textit{y,z}) and vertical (\textit{x}) polarization. The TOF-MS axis is placed perpendicular to the $1\omega$ and $2\omega/4\omega$ laser beam plane, along \textit{x}-axis. For the $E_{x\pm iy,x,z}$ configuration a tunable dual wavelength waveplate was added into the $2\omega/4\omega$ optical path after the wedges.}
    \label{fig:optics}
\end{figure}

The pulses were delivered onto a polarizing beamsplitter generating two pulses with mutually perpendicular polarizations in the two beamlines (Fig. \ref{fig:optics}). The transmitted, horizontally polarized  pulses passed through a motorized delay stage and then a half waveplate (HWP) followed by a quarter waveplate (QWP) to generate circularly polarized 1$\omega$ pulses. A 20 cm focal length lens focused the pulses through a vacuum window and into the interaction region where they crossed pulses from a second perpendicular beamline 8.5 mm from the focal position. 

In the second beamline, vertically ($E_x$) polarized 1$\omega$ pulses passed through a BBO crystal to generate second harmonic 2$\omega$ pulses with the remaining fundamental light removed by two dichroic mirrors (not shown in Fig. \ref{fig:optics}). The 2$\omega$ pulses passed through a half waveplate and then a 3~mm thick calcite crystal. The calcite generated a vertically polarized pulse 2$\omega$($E_x$) and a horizontally polarized pulse 2$\omega$($E_z$) separated in time by 2~ps. A half waveplate was used to rotate the polarization axis relative to the optical axis of the calcite to control the relative pulse energies of the two emerging pulses. 

A second BBO crystal was orientated to optimize production of the 4th harmonic  polarized in the horizontal direction 4$\omega$($E_z$) from the 2$\omega$($E_x$) pulses. The pulses then passed through a pair of thin fused silica wedges before being focused by a 25 cm focal length spherical mirror through a calcium fluoride vacuum window into the interaction region where they crossed the 1$\omega$ pulses from the other beamline 15 mm after the focal point. To produce synthetic chiral light from 2$\omega$ and 4$\omega$ pulses with parallel polarizations, $E_{x\pm iy,z,z}$, the dispersion of the wedges and vacuum window introduced a delay to the 4$\omega$($E_z$) pulses relative to the 2$\omega$($E_z$) pulses so that they were temporally overlapped when they reached the interaction region. 

To generate chiral light pulses from 2$\omega$ and 4$\omega$ pulses with perpendicular polarization, $E_{x\pm iy,x,z}$, a tunable dual wavelength waveplate (ALPHALAS GmbH) was introduced after the wedges in the 2$\omega$/4$\omega$ beamline. This was configured as a half waveplate for the 2$\omega$ pulses and a full waveplate for 4$\omega$ to orientate the 2$\omega$ polarization vertical ($E_x$) while leaving the 4$\omega$ horizontal ($E_z$). To determine the intensity of all three pulses, the average power of each frequency component was recorded and the beam profiles at the off-focus distances measured on a CCD camera. 

\subsubsection*{1.2 Time-of-Flight (ToF) Mass Spectrometer}
\begin{figure}
    \centering
    \includegraphics[width=0.6\linewidth]{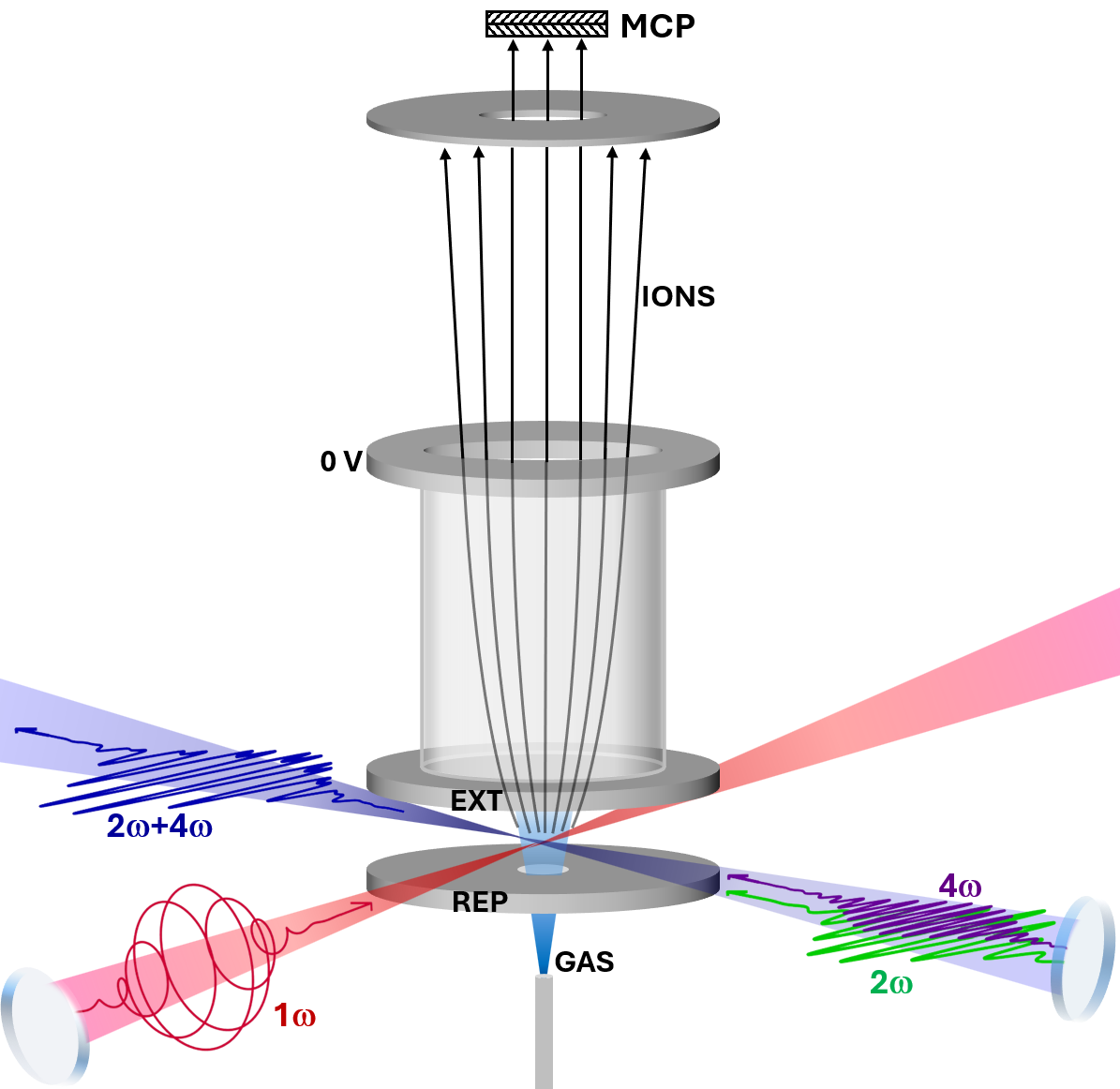}
    \caption{\textbf{The chiral ToF mass spectrometer coupled to optics which generates synthetic chiral light in the parallel 2$\omega$/4$\omega$ configuration $E_{x\pm iy,z,z}$}. A circularly polarized pulse at the fundamental frequency (1$\omega$) intersects perpendicularly with 2nd (2$\omega$) and 4th (4$\omega$) harmonic pulses to generate synthetic chiral light within a chiral molecular gas jet (GAS). The potential difference between the repeller (REP) and extraction (EXT) plates accelerates the resulting photoions into a flight tube where they are separated temporally by their mass and detected by a micro-channel plate (MCP).}
    \label{fig:ChiralTOF}
\end{figure}

The pulses from the two beamlines crossed perpendicularly on the axis of a home-built time-of-flight mass spectrometer which has been described in previous publications \cite{calegari2014,greenwood2025}. A schematic diagram of how it was merged with the synthetic chiral light optics is shown in Fig. \ref{fig:ChiralTOF}. The laser pulses entered between a repeller and an extraction plate with potentials 4000~V and 2500~V respectively. The extraction plate was connected to one end of a cylindrical tube with a resistive coating on its inner surface, while the other end was grounded. The electric field generated by these potentials was designed to only transmit ions created close to the ToF axis (within an estimated $\sim$500$ \mu$m range) onto a Photonis MiniTOF micro-channel plate (MCP) detector. This suppressed ions which were not produced close to the intersection of the beams, allowing a drastic background reduction. Pulses generated from the detector passed through an amplifier-discriminator (ET Enterprises AD8) to create 17~ns TTL pulses which were directed into 4 separate digital counters. Each counter was gated with a time delayed pulse corresponding to a specific ion in the mass spectrum. 

The molecular samples were introduced into the interaction region by an effusive gas jet which passed through an aperture in the repeller plate. The base vacuum inside the mass spectrometer was $10^{-8}$~mbar and rose to around $3\times10^{-5}$~mbar during data acquisition. 

\begin{figure} 
	\centering
	\includegraphics[width=\textwidth]{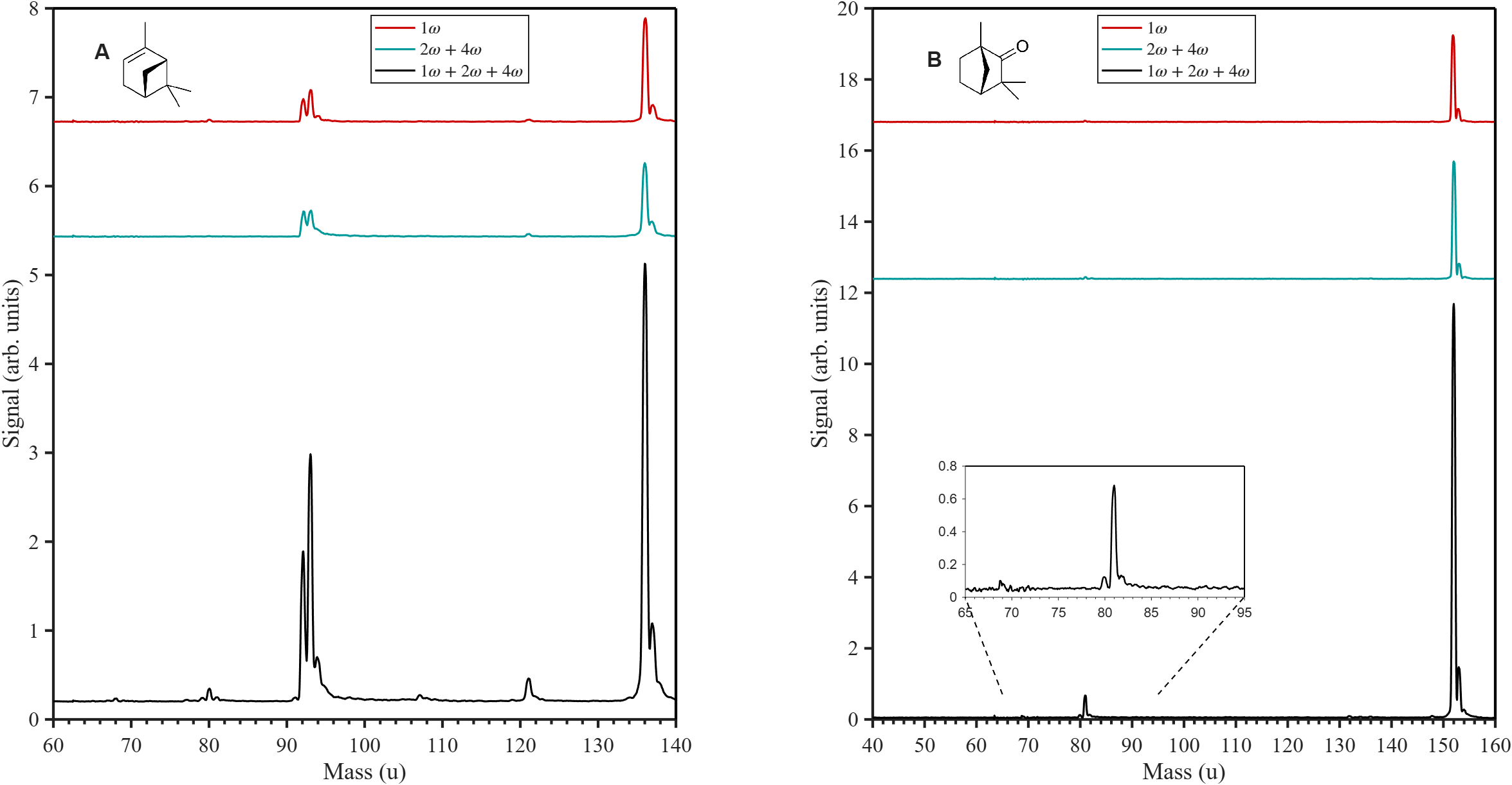}   
    \caption{ \textbf{Time-of-Flight mass spectra for $\alpha$-pinene (A) and fenchone (B)}.  Three spectra are shown for each molecule corresponding to ionization by the 1$\omega$ pulses only (red), the 2$\omega$/4$\omega$ pulses (blue), and all three pulses spatially and temporally overlapped (black). The parent molecular ion (masses 136u and 152u respectively) dominates each spectrum and it is only when all three colors are present that the fragment ions become significant.}  
	\label{fig:TOF} 
\end{figure}

Figure \ref{fig:TOF} shows time-of-flight mass spectra for $\alpha$-pinene and fenchone acquired for the 1$\omega$ pulses only, the 2$\omega$ and 4$\omega$ pulses together, and with all three pulses. The low intensity of the individual pulses results in the parent molecular ion dominating the spectrum, with small amounts of fragment ions produced. When all three pulses are spatially and temporally overlapped, there is a strong enhancement in the number of ions produced, particularly the fragment ions. For fenchone the main fragments at masses 69 and 81 are complementary ions generated from cleavage of the cyclic ring beside the carbonyl and methyl groups along with an H atom transfer. For $\alpha$-pinene the fragment ions at 92/93 (produced from ring opening followed by loss of a $C_{3}H_{8}$ or $C_{3}H_{7}$ radical) were almost as strong as the parent ion with weaker contributions at 121 (loss of a methyl group) and 80.

\subsubsection*{1.3 Photoion Circular Dichroism (PICD) in fenchone and $\alpha$-pinene}
The ToF mass spectrometer has previously been used to study PICD from 1-color and 2-color multiphoton ionization using a laser with a 2nd harmonic wavelength of 520 nm and 4th harmonic at 260 nm \cite{greenwood2025}. The PICD is defined as the normalized difference in the ion yields obtained from the laser pulses being left- or right-circularly polarized, $I_L$ and $I_R$.
\begin{equation}
\textrm{PICD}=2\frac{I_L-I_R}{I_L+I_R}
\end{equation}

The PICD values obtained from these studies in $\alpha$-pinene and fenchone are presented in Table \ref{tab:PICD}. The fenchone values were published in Greenwood et al. \cite{greenwood2025}, while the results for $\alpha$-pinene were not previously published. The difference in the ion yields between left- and right-circularly polarizations is very small (PICD $<$0.2\% in fenchone, $<0.3\%$ in $\alpha$-pinene), irrespective of the color used. This demonstrates that conventional circular dichroism arising from non-electric dipole effects in these molecules is more than an order of magnitude less than the SCD measured in the present study.  

\begin{table} 
	\centering
	\caption{\textbf{One-Color Photoion Circular Dichroism (PICD) in (+)-$\alpha$-pinene and (+)-fenchone.}
		One-color PICD values measured for different mass ions (parent molecular ion mass 152 for fenchone and 136 for $\alpha$-pinene) at wavelengths 260 nm and 520 nm. Uncertainties are one standard error in the mean. }
	\label{tab:PICD} 
	
	\begin{tabular}{ccc} 
		\\
		\hline
        & (+)-fenchone & \\
		Mass & 520 nm & 260 nm \\
		 \hline
		152 & -0.17$\pm$0.02\% & 0.15$\pm$0.02\% \\
		81 & 0.23$\pm$0.06\% & 0.30$\pm$0.08\% \\
		69 & -0.05$\pm$0.16\%  & 0.31$\pm$0.07\% \\
		\hline
	\end{tabular}
    \\
    \begin{tabular}{ccc} 
		\\
		\hline
        & (+)-$\alpha$-pinene & \\
		Mass & 520 nm & 260 nm\\
		 \hline
		136 & -0.21$\pm$0.03\% & 0.00$\pm$0.02\%  \\
		121 &  0.04$\pm$0.07\% & 0.05$\pm$0.12\%  \\
		106 & 0.16$\pm$0.08\% & -0.04$\pm$0.10\%  \\
		92/93 & -0.23$\pm$0.04\%  & 0.05$\pm$0.08\%  \\
		\hline
	\end{tabular}
\end{table}

\subsubsection*{1.4 2$\omega$/4$\omega$ Phase $\Phi_{4\omega}$ – Calibration and Stability }
\begin{figure}
	    \centering
	    \includegraphics[width=0.5\linewidth]{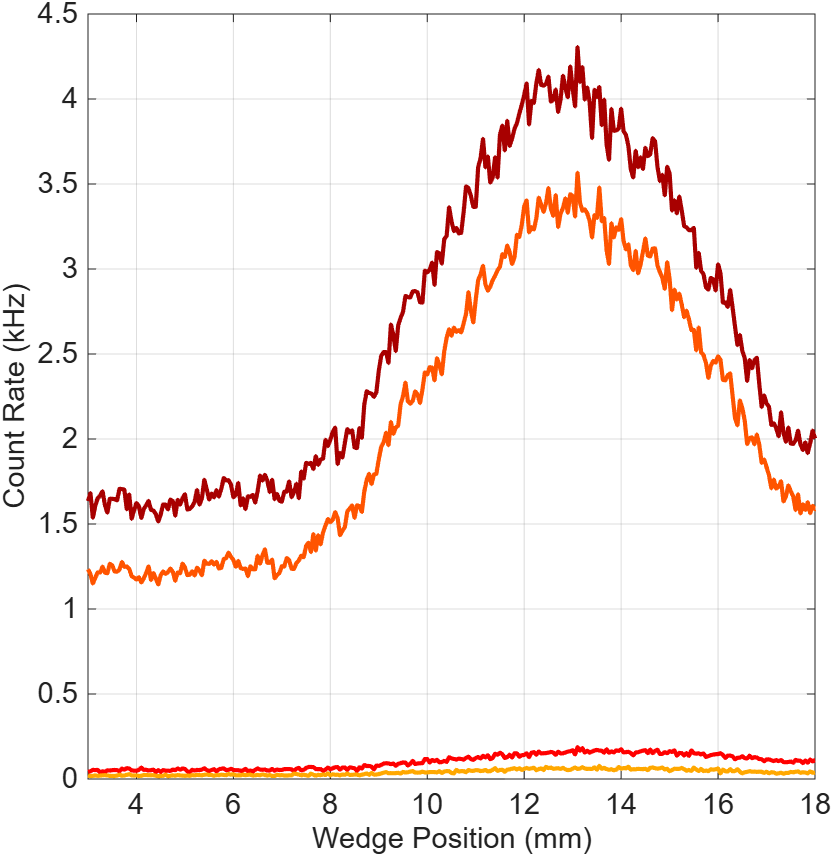}
	    \caption{Ion count rate generated from (+)-$\alpha$-pinene using pulses only from the 2$\omega$/4$\omega$ beamline with their polarizations in a perpendicular configuration as a function of the silica wedge position. The curves are for ion yields with masses 136, 92/93, 121 and 80 from top to bottom.}
  	    \label{fig:wedge_max}
\end{figure}

In Fig. \ref{fig:wedge_max} the ion yield count rate for $\alpha$-pinene generated by pulses from the 2$\omega$/4$\omega$ beamline with orthogonal polarizations is displayed as a function of the fused silica wedge position. The wedges finely control the temporal separation between the two laser frequencies. The peak in the count rate corresponds to the optimum temporal overlap between the 2$\omega$($E_x$) and 4$\omega$($E_z$) pulses. 

To determine the phase $\Phi_{4\omega}$ stability and calibration against the wedge translation, a stereo-electron detector was adapted to measure left-right asymmetry in the photoelectron emission (previously it had been used to measure the forward-back asymmetry from Photoelectron Circular Dichroism \cite{miles2017,greenwood2023}). For this measurement, the 2$\omega$ and 4$\omega$ pulses were configured with their polarizations parallel in the horizontal direction. A flip mirror was used to pick off the pulses and send them into this instrument using a similar focusing arrangement to that used for the main instrument (Fig. \ref{fig:optics}). When the polarizations of the pulses were parallel, the direction of maximum electric field of the combined pulses varied between left and right as $\Phi_{4\omega}$ changed. This asymmetry in the field induced an asymmetry in the left vs right photoelectron emission.

\begin{figure} 
	\centering
	\includegraphics[width=\textwidth]{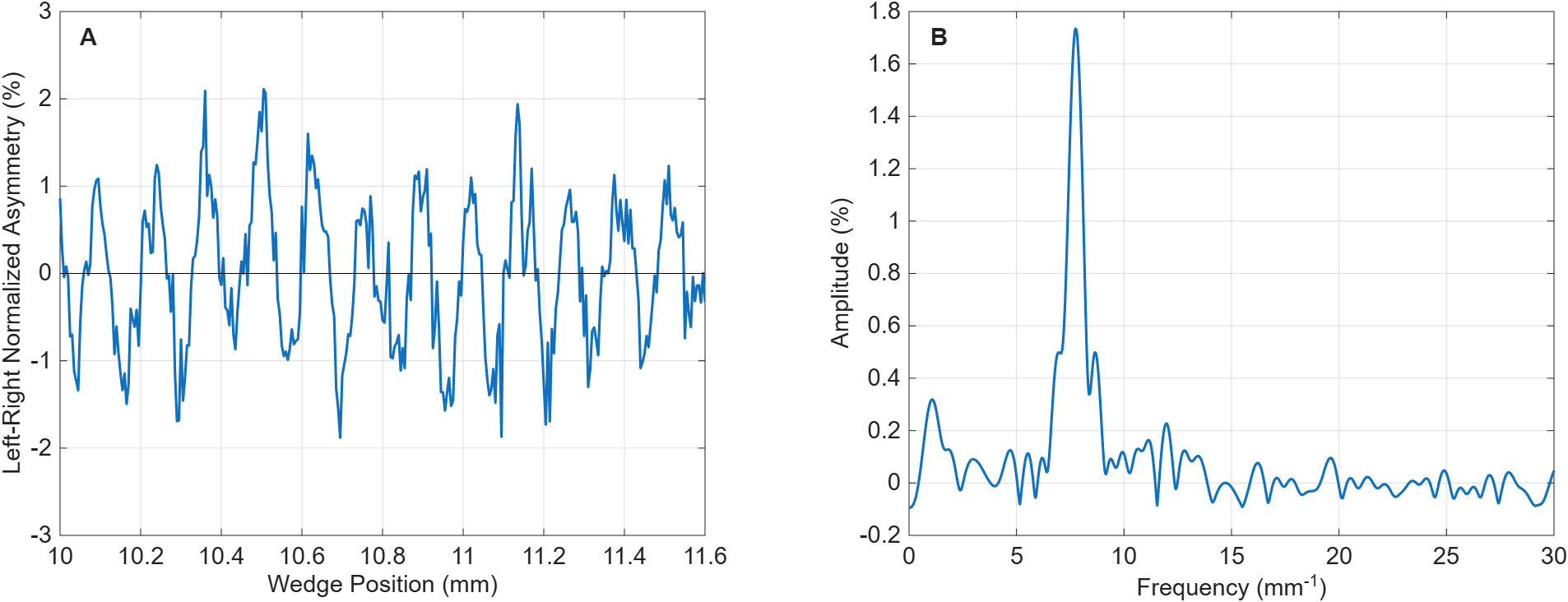} 

	\caption{\textbf{A}. Normalized left-right photoemission asymmetry generated from the 2$\omega$/4$\omega$ beamline as a function of the silica wedge position. \textbf{B}. Fast Fourier Transform of A.} 
\label{fig:wedge_phase} 
\end{figure}

The normalized difference in the left-to-right ionization rate of fenchone as a function of wedge position, as presented in Fig. \ref{fig:wedge_phase}, shows oscillations with a period of 0.129 $\pm$ 0.001 mm. As this asymmetry is expected to vary with the 4$\omega$ time period (0.86~fs), a 1~mm increase in the wedge position corresponded to a 6.6~fs time delay of the 4$\omega$ pulse relative to 2$\omega$. Small shifts in the peaks relative to the average period were observed with a standard deviation of $\pm$0.005 mm giving a stability in $\Phi_{4\omega}$ of $\pm$0.075$\pi$ (33 attoseconds). Over acquisition times of several hours, larger phase drifts were sometimes observed in the SCD measurements. These most likely originate from temperature variations in the room, as well as from heating of optical elements by the high power beam -- the average power of the laser beam impinging on the beamsplitter shown in Fig \ref{fig:optics} was 30 W. Due to different optical paths and optical elements, the absolute value of $\Phi_{4\omega}$ measured with the stereo electron spectrometer cannot be directly correlated with the SCD measurements in the ToF instrument.

\subsubsection*{1.5 Contribution of 3-Color Ionization to the Total Count Rate}
\begin{figure}
	\centering
	\includegraphics[width=0.5\textwidth]{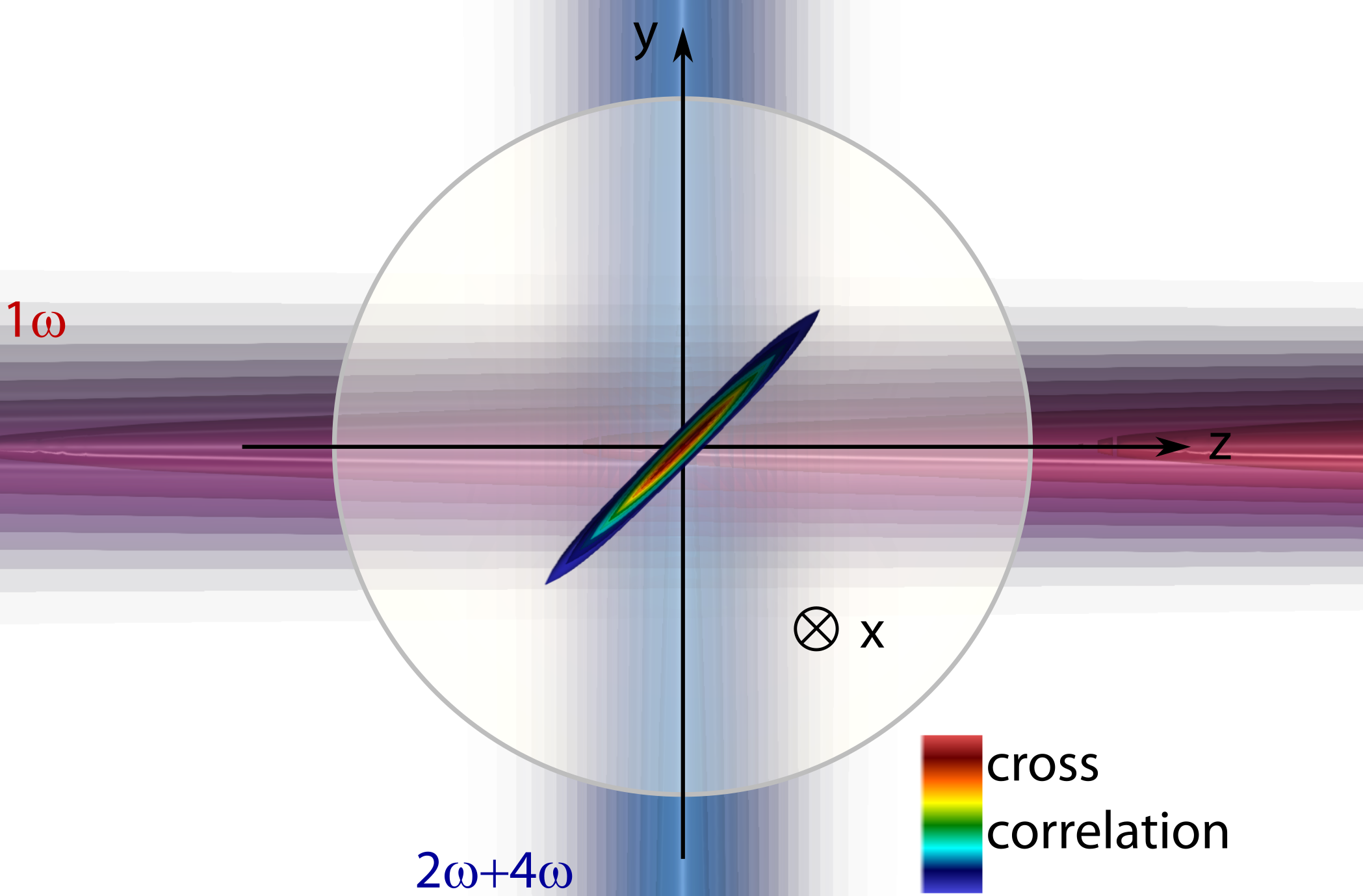} 
	\caption{Schematic representation of the interaction region. The circle, with 500~$\mu$m diameter, represents the estimated collection area of the mass-spectrometer. The colored diagonal area is the cross-correlation of two intersecting 100 fs pulses.} 
\label{fig:crossco} 
\end{figure}
The fraction of the count rate for ion $i$ produced by absorption from all of the laser pulses (at their temporal overlap) is $F^i$. The background count rate $(1-F^i)$ arises from ionization involving only one or two of the three colors. The non-collinear geometry of the interaction reduces the overlap region of the intersecting beams to a $\sim(100)^3~\mu$m$^3$ volume. Furthermore, the short duration of the laser pulses restricts their temporal overlap to a slice of this volume. A schematic of this overlap region is shown in Fig. \ref{fig:crossco}, which depicts the intersecting 1$\omega$  and 2$\omega$/4$\omega$ laser beams. The diagonal colored area represents the cross correlation between two intersecting 100~fs pulses. Its thickness, determined by the pulse duration, is about 50~$\mu$m, and its length, set by the waist of the intersecting beams, is in the 140~$\mu$m range.  To enhance the values of $F^i$, relatively low laser intensities were used for the individual beams while still achieving a total count rate of around 50 kHz. Most of the ionization was generated outside the volume where the pulses are spatially and temporally overlapped. The design of the ToF spectrometer, which only detects the ions produced within a $\sim$500~$\mu$m disk, enables the background ionization originating from the individual beams to be lowered. It might be beneficial for future setups to use picosecond pulses to increase the 3-color overlap volume, particularly if chiral processing with more powerful, larger diameter lasers is desired. 

\begin{figure} 
	\centering
	\includegraphics[width=0.95\textwidth]{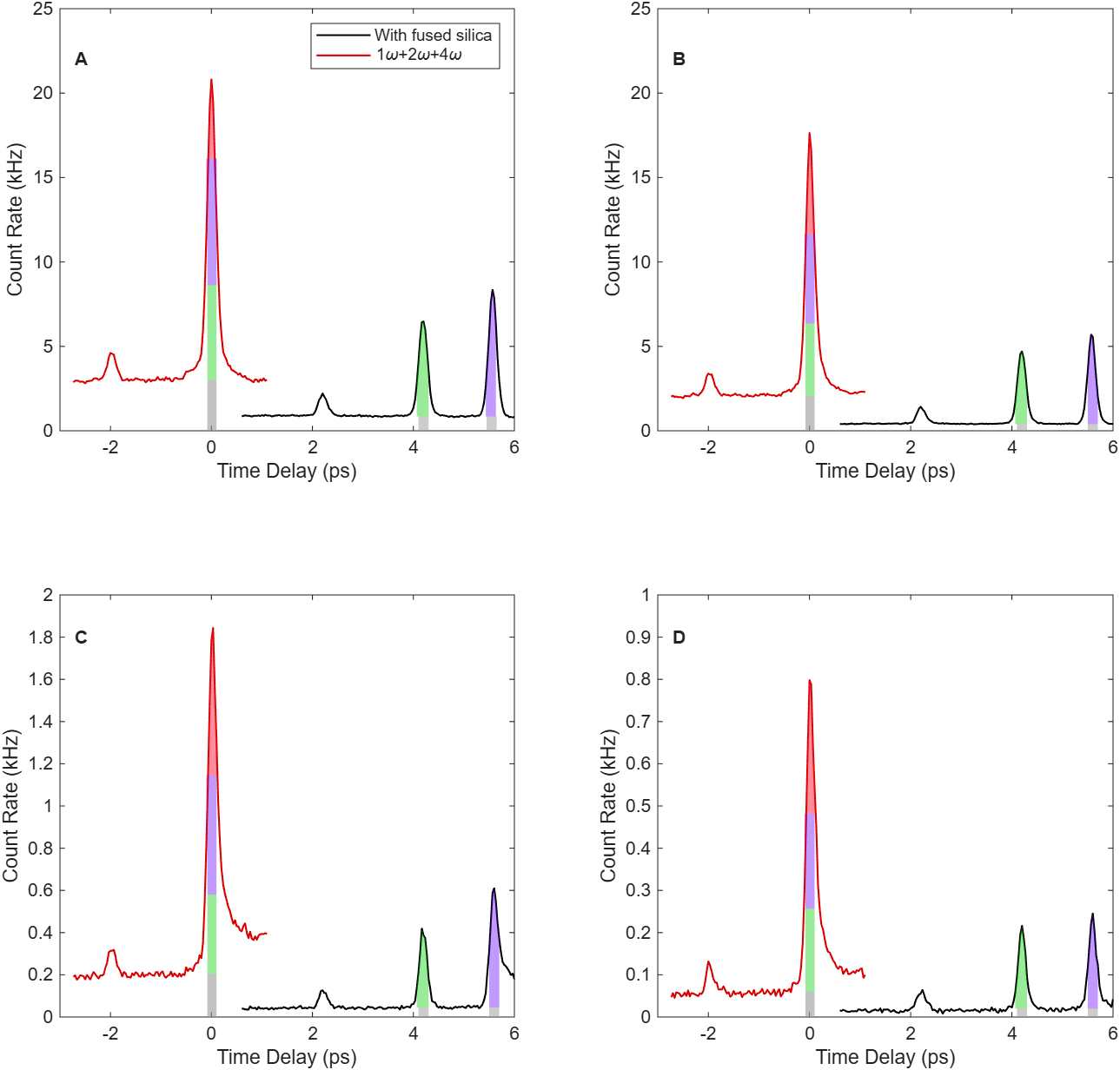} 

	\caption{ \textbf{Delay scans for $\alpha$-pinene used to calculate \textit{F}.}  A,B,C,D show the count rates for the parent (136u) and fragment ion channels (92/93u, 121u, 80u) respectively as a function of the delay of the 1$\omega$ pulses relative to the 2$\omega$/4$\omega$ beamline for the $E_{x+iy,x,z}$ configuration. The black and red lines are scans with and without a 5 mm fused silica plate inserted in the 2$\omega$/4$\omega$ beamline to temporally separate the 2$\omega$ and 4$\omega$ pulses. The shaded regions show contributions due to the cross correlation of the 1$\omega$ pulse with the 2$\omega$ (green shading) and 4$\omega$ (lilac shading) pulses. The background contributions to the count rates are shown by grey shading. Without the silica plate at zero time delay there is large signal enhancement from the overlap of all three frequencies, with the red shading under the peak showing the fraction $F^i$ of the count rate which arises from ionization involving at least one photon from all three colors. See text for further details.}  
	\label{fig:delay} 
\end{figure}

To obtain temporal overlap between the 1$\omega$ and the 2$\omega$/4$\omega$ pulses, the maximum in the ionization count rate was recorded as the delay between the pulses was incremented using the delay stage in the 1$\omega$ beamline. Fig. \ref{fig:delay} presents data for $\alpha$-pinene using synthetic chiral pulses with 2$\omega$/4$\omega$ in the perpendicular polarization configuration $E_{x\pm iy,x,z}$. Negative delays correspond to the 1$\omega$ pulse arriving at the interaction region before the 2$\omega$/4$\omega$ pulses. The peak at a negative delay (red line) corresponds to temporal overlap of 1$\omega$ with the 2$\omega$($E_z$) pulses used to generate the 4$\omega$ pulses. The enhancement at zero delay is due to the spatial and temporal overlap of all three 1$\omega$, 2$\omega$($E_x$) and 4$\omega$($E_z$) pulses needed to synthesize the locally and globally chiral light. This cross-correlation peak is slightly asymmetric due to some dynamics in excited states populated by the 2$\omega$($E_x$)/4$\omega$($E_z$) pulses being probed in the ionization step by a delayed 1$\omega$ pulse. This is more evident for the fragment ion (121u, 80u) production at positive delays. More significant dynamics were obtained for delay scans in fenchone (see \cite{comby2016,blanchet2021} for previous discussion of this dynamics).  

When there is no overlap between the pulses, there is a background count rate due to ionization from the 1$\omega$ pulse only and from a combination of the 2$\omega$($E_z$) and 2$\omega$($E_x$)+4$\omega$($E_z$) pulses (grey shading). However, to extract the fraction $F^i$ arising from ionization involving at least one photon from each of the three different colors, the additional background due to 2-color ionization from 1$\omega$+2$\omega$($E_x$) and 1$\omega$+4$\omega$($E_z$) must be estimated. For this, a 5 mm fused silica plate was introduced in the beamline to temporally separate the 2$\omega$($E_x$) and 4$\omega$($E_z$) pulses and another delay scan taken. 

Fig.  \ref{fig:delay} shows an additional plot (black line) with three peaks corresponding to overlap of the 1$\omega$ with 2$\omega$($E_z$), 2$\omega$($E_x$) and 4$\omega$($E_z$) pulses at increasing delays respectively. The enhancement found in the two latter peaks (green and lilac shading) were then used to determine rates arising from the 1$\omega$+2$\omega$($E_x$)  and 1$\omega$+4$\omega$($E_z$) ionization processes. The relative contributions to the count rate of the zero delay peak are displayed as shaded bars under the peak with the 3-color contribution shaded red. Table \ref{tab:Fraction F} presents the fraction $F^i$ of the count rate corresponding to the ionization involving all three colors for the two molecules and the two different chiral light configurations used in the measurements.

\begin{table} 
	\centering
	\caption{\textbf{Fraction $F^i$ of Count Rates due to 3-Color Ionization.} Determination of the fraction of the count rate obtained when all 3 pulses are overlapped which is due to absorption of at least one photon from each color, as demonstrated in Fig.~\ref{fig:delay}. Values of $F^i$ for $\alpha$-pinene and fenchone were obtained for 2$\omega$ and 4$\omega$ polarizations in the parallel ($E_{x\pm iy,z,z}$) and perpendicular ($E_{x\pm iy,x,z}$) configurations. The uncertainty $\Delta F^i$ was estimated at 90\% confidence.\\}
	
    \label{tab:Fraction F} 

    \textbf{$\alpha$-pinene}\\
	\begin{tabular}{ccccccc}
        \hline 
         & & $\mathbf{E_{x\pm iy,z,z}}$ & & & $\mathbf{E_{x\pm iy,x,z}}$ &  \\
		  Mass & & $\mathbf{F^i}$ & $\Delta F^i$ & & $\mathbf{F^i}$ & $\Delta$\textit{F} \\
        \hline
		136 & & \textbf{0.14} & 0.04 & & \textbf{0.23} & 0.04  \\
        121 & & \textbf{0.33} & 0.11 & & \textbf{0.38} & 0.06\\
        92,93 & & \textbf{0.25} & 0.04 & & \textbf{0.34} & 0.04   \\
        80 & & \textbf{0.36} & 0.10 & & \textbf{0.39} & 0.06   \\
		  \hline
          \\
	\end{tabular}
    
    \textbf{Fenchone}\\
	\begin{tabular}{ccccccc}
        \hline 
         & & $\mathbf{E_{x\pm iy,z,z}}$ & & & $\mathbf{E_{x\pm iy,x,z}}$ &  \\
		  Mass & & $\mathbf{F^i}$ & $\Delta F^i$ & & $\mathbf{F^i}$ & $\Delta F^i$ \\
        \hline
		152 & & \textbf{0.23}& 0.03 & & \textbf{0.19} & 0.03  \\
        81 & & \textbf{0.48} & 0.04 & & \textbf{0.42} & 0.05\\
        69 & & \textbf{0.51} & 0.07 & & \textbf{0.36} & 0.18   \\
		  \hline
          \\
	\end{tabular}
\end{table}

\subsubsection*{1.6 Data Acquisition and Samples}
Having established the zero time delay position, the number of ion counts were recorded for a 10~s duration with the 1$\omega$ pulse left circularly polarized. By rotating the half waveplate by 45°, the ion counts for right-circularly polarized light was also measured over the same period. The position of one of the fused silica wedges was then incremented a distance of 9.29 $\mu$m to change the relative phase $\Phi_{4\omega}$ between the 2$\omega$ and 4$\omega$ pulses and the data acquired for the two 1$\omega$ polarization states again. This continued until $\Phi_{4\omega}$ was shifted by a total of 12$\pi$ in 84 steps and the whole process repeated for at least 20 cycles, taking about 10 hours. The chiral sample was then switched and the opposite enantiomer measured under the same conditions. To ensure there was no cross-contamination, the previous sample was isolated with a valve and the gas line pumped out until the ion count rate had reduced by more than three orders of magnitude. It was only then that gas from the opposite enantiomer was introduced into the system. 

The chiral chemicals were purchased from Sigma-Aldrich and used without any further purification. The quoted chemical and enantiomeric purities of the samples are given in Table \ref{tab:samples}.

\begin{table} 
	\centering
	\caption{\textbf{Samples: chemical and chiral purities.} Sigma-Aldrich product number, quoted chemical purity and optical activity (O.A.) and estimated enantiomeric excess (e.e.) of the four samples used for the measurements. Values for e.e. were estimated assuming the absolute optical rotation for enantiopure samples of $\alpha$-pinene and fenchone are 45$^\circ$ and 60$^\circ$ respectively.}
	
    \label{tab:samples} 

    \textbf{$\alpha$-pinene}\\
	\begin{tabular}{ccccc}
        \hline 
        \textbf{Enantiomer} & $\mathbf{No.}$ & $\mathbf{Purity}$ & $\mathbf{O.A.}$ & $\mathbf{e.e}$  \\
         \hline
		  (+) & W290238 & $>$96.5\% & $[\alpha]20/D +42\pm3^\circ$,neat & 93$\%$ \\
		(-) & W290203 & $>$98\% & $[\alpha]20/D +41^\circ$,neat & 91$\%$ \\
		  \hline
          \\
	\end{tabular}
    
    \textbf{Fenchone}\\
	\begin{tabular}{ccccc}
        \hline 
        \textbf{Enantiomer} & $\mathbf{No.}$ & $\mathbf{Purity}$ & $\mathbf{O.A.}$ & $\mathbf{e.e}$  \\
         \hline
		  (+) & 46210-F & $-$ & $[\alpha]20/D +60\pm3^\circ$,neat & $>$98$\%$ \\
		(-) & 196436 & $>$98\% & $[\alpha]24/D +50.5^\circ$,neat & 84$\%$ \\
		  \hline

	\end{tabular}
\end{table}

\subsubsection*{1.7 Analysis of the Synthetic Chiral Dichroism signal}
The data obtained at each wedge position and fundamental beam handedness are summed, and the signals from opposite handedness are subtracted to obtain the SCD. The resulting signal shows 6 consecutive oscillations, associated with the $12\pi$ scan of $\Phi_{4\omega}$. We calculate the average value of the SCD: 
\begin{equation}
SCD(\Phi_{4\omega})=\frac{\sum_{k=1}^{6}{SCD(\Phi_{4\omega}+2k\pi)}}{6}
\end{equation}
and subtract the signals obtained at $\pi$-shifted values of $\Phi_{4\omega}$:
\begin{equation}
SCD(\Phi_{4\omega})=\frac{1}{2}\left[SCD(\Phi_{4\omega})-SCD(\Phi_{4\omega}+\pi)\right]
\end{equation}

The signals from individual enantiomers are analyzed independently. The 90$\%$ confidence interval is obtained from the standard deviation of the 6 measurements: 
\begin{equation}
\delta SCD(\Phi_{4\omega})=t_{(6,90\%)}\frac{\sigma(SCD)}{\sqrt{6}}
\end{equation}
where $t_{(6,90\%)}\approx1.9$ is the corresponding Student coefficient. 

A sinusoidal fit of the resulting signal is performed to extract the amplitude and phase of the oscillations. The error bars on the fit are obtained by repeating the fitting procedure 200 times randomly picking each data point within the confidence interval of the SCD measurement.
The results are in Tables \ref{tab:SCDpinene} and \ref{tab:SCDfenchone}.

\begin{table}[]
	\centering
	\caption{\textbf{Amplitude (in \%) and phase (in rad) of the SCD modulations in $\alpha$-pinene.}
		The error bars correspond to the 95$\%$ confidence interval of the sinusoidal fit. Results demonstrating a statistically significant phase shift of the SCD oscillations in both enantiomers with respect to the parent signal are in bold.}
	\label{tab:SCDpinene} 
\begin{tabular}{|l|llll|llll|}
\cline{1-9}
Configuration & \multicolumn{4}{l|}{ $E_{x+iy,x,z}$}                          & \multicolumn{4}{l|}{$E_{x+iy,z,z}$}                            \\ \cline{1-9}
Mass   & \multicolumn{1}{l|}{136} & \multicolumn{1}{l|}{121} & \multicolumn{1}{l|}{92}  & 80 & \multicolumn{1}{l|}{136} & \multicolumn{1}{l|}{121} & \multicolumn{1}{l|}{92}  & 80  \\ \cline{1-9}
SCD$^{(+)}$ (\%) &  \multicolumn{1}{l|}{4.1$\pm$0.2} & \multicolumn{1}{l|}{0.9$\pm$0.1} & \multicolumn{1}{l|}{1.5$\pm$0.1} & 1.1$\pm$0.1  & \multicolumn{1}{l|}{2.2$\pm$0.2}    & \multicolumn{1}{l|}{1$\pm$1}  & \multicolumn{1}{l|}{0.9$\pm$0.1} & 0.7$\pm$0.1    \\ \cline{1-9}
SCD$^{(-)}$ (\%) &  \multicolumn{1}{l|}{2.6$\pm$0.2} & \multicolumn{1}{l|}{0.9$\pm$0.1} & \multicolumn{1}{l|}{1.0$\pm$0.1} & 1.2$\pm$0.1  & \multicolumn{1}{l|}{1.2$\pm$0.2}    & \multicolumn{1}{l|}{0.3$\pm$0.2} & \multicolumn{1}{l|}{0.4$\pm$0.1}  & 0.3$\pm$0.1    \\ \cline{1-9}
$\Phi_{4\omega}^{(+)}-\Phi_{4\omega}^{(+),136}$& \multicolumn{1}{l|}{0.0$\pm$0.1} & \multicolumn{1}{l|}{-0.2$\pm$0.2} & \multicolumn{1}{l|}{-0.1$\pm$0.1} & -0.4$\pm$0.2  & \multicolumn{1}{l|}{0.0$\pm$0.2}    & \multicolumn{1}{l|}{3$\pm$3} & \multicolumn{1}{l|}{\textbf{1.0$\pm$0.2}}  &\textbf{1.4$\pm$0.4}    \\ \cline{1-9}
$\Phi_{4\omega}^{(-)}-\Phi_{4\omega}^{(-),136}$& \multicolumn{1}{l|}{0.0$\pm$0.1} & \multicolumn{1}{l|}{-0.2$\pm$0.2} & \multicolumn{1}{l|}{-0.1$\pm$0.1} & 0.0$\pm$0.2  & \multicolumn{1}{l|}{0.0$\pm$0.2}    & \multicolumn{1}{l|}{0.9$\pm$0.2} & \multicolumn{1}{l|}{\textbf{0.5$\pm$0.1}}  & \textbf{1.1$\pm$0.2   } \\ \cline{1-9}
$\Phi_{4\omega}^{(-)}-\Phi_{4\omega}^{(+)}$& \multicolumn{1}{l|}{3.6$\pm$0.2} & \multicolumn{1}{l|}{3.7$\pm$0.4} & \multicolumn{1}{l|}{3.6$\pm$0.2} & 4.0$\pm$0.4  & \multicolumn{1}{l|}{2.9$\pm$0.4}    & \multicolumn{1}{l|}{2$\pm$3} & \multicolumn{1}{l|}{2.4$\pm$0.3}  & 2.6$\pm$0.6    \\ \cline{1-9}
\end{tabular}
\end{table}

\begin{table}[]
	\centering
	\caption{\textbf{Amplitude (in \%) and phase (in rad) of the SCD modulations in fenchone.}
		The error bars correspond to the 95$\%$ confidence interval of the sinusoidal fit. The results demonstrating a statistically significant phase shift of the SCD oscillations in both enantiomers with respect to the parent signal are in bold.}
	\label{tab:SCDfenchone} 
\begin{tabular}{|l|lll|lll|}
\cline{1-7}
Configuration & \multicolumn{3}{l|}{ $E_{x+iy,x,z}$}                          & \multicolumn{3}{l|}{$E_{x+iy,z,z}$}                            \\ \cline{1-7}
Mass   & \multicolumn{1}{l|}{152} & \multicolumn{1}{l|}{81} & 69 & \multicolumn{1}{l|}{152} & \multicolumn{1}{l|}{81} & 69  \\ \cline{1-7}
SCD$^{(+)}$ (\%) &  \multicolumn{1}{l|}{2.2$\pm$0.2} & \multicolumn{1}{l|}{1.3$\pm$0.1} & 1.3$\pm$0.2  & \multicolumn{1}{l|}{1.3$\pm$0.1}    & \multicolumn{1}{l|}{0.7$\pm$0.1}   & 0.9$\pm$0.1    \\ \cline{1-7}
SCD$^{(-)}$ (\%) &  \multicolumn{1}{l|}{1.4$\pm$0.1} & \multicolumn{1}{l|}{0.8$\pm$0.1} & 0.9$\pm$0.1  & \multicolumn{1}{l|}{1.1$\pm$0.1}    & \multicolumn{1}{l|}{1.0$\pm$0.1}   & 1$\pm$0.1    \\ \cline{1-7}
$\Phi_{4\omega}^{(+)}-\Phi_{4\omega}^{(+),152}$ (rad) & \multicolumn{1}{l|}{0.0$\pm$0.1}    & \multicolumn{1}{l|}{\textbf{0.3$\pm$0.1}}   &  0.2$\pm$0.3  & \multicolumn{1}{l|}{0.0$\pm$0.2}    & \multicolumn{1}{l|}{-0.3$\pm$0.3}   &  \textbf{0.4$\pm$0.3}   \\ \cline{1-7}
$\Phi_{4\omega}^{(-)}-\Phi_{4\omega}^{(-),152}$  (rad) & \multicolumn{1}{l|}{0.0$\pm$0.2}    & \multicolumn{1}{l|}{\textbf{0.4$\pm$0.1}}   &  0.2$\pm$0.3  & \multicolumn{1}{l|}{0.0$\pm$0.1}    & \multicolumn{1}{l|}{0.1$\pm$0.1}   &   \textbf{0.6$\pm$0.1}  \\ \cline{1-7}
$\Phi_{4\omega}^{(-)}-\Phi_{4\omega}^{(+)}$  (rad) & \multicolumn{1}{l|}{3.9$\pm$0.3}    & \multicolumn{1}{l|}{3.9$\pm$0.2}   &  3.8$\pm$0.6  & \multicolumn{1}{l|}{2.8$\pm$0.3}    & \multicolumn{1}{l|}{3.2$\pm$0.4}   &   2.9$\pm$0.4  \\ \cline{1-7}
\end{tabular}
\end{table}

\subsubsection*{1.8 Analysis of the Branching Ratios}
\begin{figure} 
	\centering
	\includegraphics[width=0.9\textwidth]{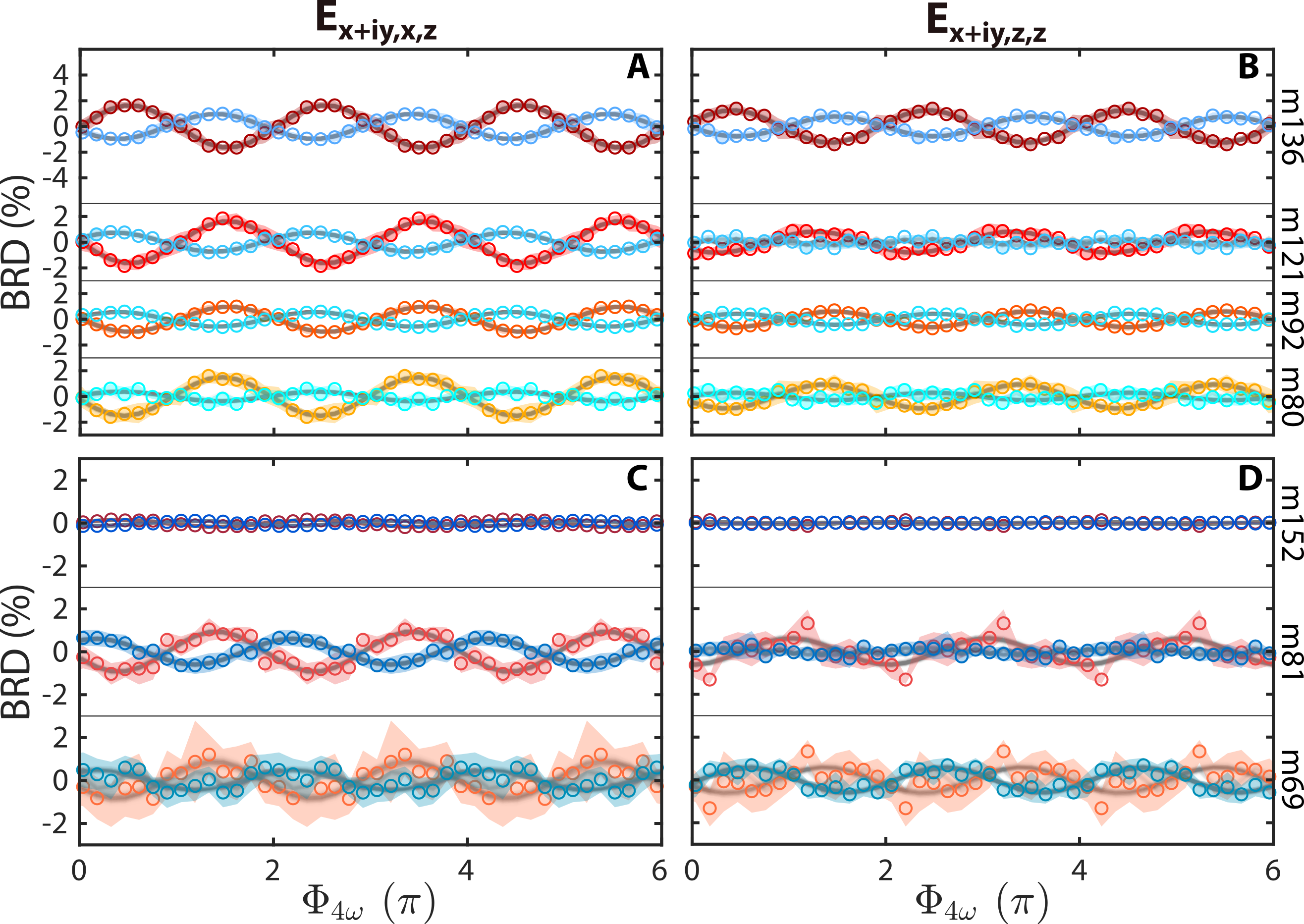} 
	\caption{\textbf{Control of branching ratios in the photoionization of $\alpha$-pinene and fenchone.}
   \textbf{(A-B)} Measured branching ratio dichroism (BRD) in (+)-$\alpha$-pinene (red to orange) and (-)-$\alpha$-pinene (blue to green) as a function of the shape of the synthetic chiral light field $E_{x\pm iy,x,z}$ (A) and $E_{x\pm iy,z,z}$ (B), controlled by scanning the phase of the fourth harmonic beam with respect to the second harmonic beam. The different lines correspond to different ions, with masses 136 (parent), and 121, 92/93, 80 (fragments) from top to bottom. The error shaded areas represent the 90$\%$ confidence interval. The semi-transparent gray lines represents 200 sinusoidal fits of sets of data randomly picked within the confidence interval. \textbf{(C-D)}  Same measurements in (+)-fenchone (red to orange) and (-)-fenchone (dark to light blue). The different lines correspond to ions with m152 (parent), m81 and m69 (fragments) from top to bottom.
		}
	\label{figBRD} 
\end{figure}

In order to highlight how selective enantio-dependent photolysis can be achieved, we calculated the branching-ratios for each ion $i$ relative to the total ion yield, $B^i=\frac{S^i}{\sum{S^i}}$, and extracted the branching-ratio dichroism: (Fig.   \ref{figBRD}).
\begin{equation}
\textrm{BRD}^i=2\frac{B_L^i-B_R^i}{B_L^i+B_R^i}
\end{equation}
This normalisation process has the advantage of reducing temporal fluctuations in the signal which arise due to variations in target gas density, laser power or beam stability, enabling the uncertainties to be reduced. This shows that branching ratios can vary by a few percent of their values, indicating that different quantum pathways are being probed and controlled by the synthetic chiral light interaction to produce different bond breakages. 

\subsubsection*{2. THEORETICAL METHODS}

\subsubsection*{2.1 Characterization of the field chirality and its coupling to chiral matter}
The chirality of the field can be quantified directly in terms of its interaction with the randomly oriented molecular sample.
For the polarizations used in this work, the simplest mechanism leading to enantio-selective population transfer corresponds to interference between the two pathways shown in Fig. \ref{fig4}A of the main text, connecting the ground and the 3s states. 
For a given polarization configuration, the difference in orientation-averaged populations between left ($-$) and right ($+$) enantiomers can be written as \cite{ordonez2025a}
\begin{equation}
P^{(+)} - P^{(-)} = \Re\{\vec{g}\cdot\vec{h}\},
\label{eq:gh}
\end{equation}
where $\vec{g}=\vec{g}^{(+)}=-\vec{g}^{(-)}$ is a complex-valued molecular pseudovector depending on transition electric dipoles and detunings [see Eq. (27) in Ref. \cite{ordonez2025a} for its explicit expression], while $\vec{h}$ is a field pseudovector (chiral correlation function) describing the chiral properties of the light, given by
\begin{equation}
\vec{h} \equiv \left(\begin{array}{c}
\lbrack \vec{E}_{2\omega} \cdot ( \vec{E}_{\omega} \times \vec{E}_{\omega}^{*} ) \rbrack ( \vec{E}_{2\omega}\cdot\vec{E}_{4\omega}^{*} )\\
\lbrack \vec{E}_{2\omega} \cdot ( \vec{E}_{\omega} \times \vec{E}_{4\omega}^{*} ) \rbrack ( \vec{E}_{2\omega}\cdot\vec{E}_{\omega}^{*} )\\ 
\lbrack \vec{E}_{2\omega} \cdot ( \vec{E}_{\omega}^{*} \times \vec{E}_{4\omega}^{*} ) \rbrack ( \vec{E}_{2\omega}\cdot\vec{E}_{\omega} )
\end{array}\right),\label{eq:h5}
\end{equation}
where $\vec{E}_{\Omega}$ is the Fourier amplitude of frequency $\Omega=\omega,2\omega,4\omega$, and its direction encodes the corresponding polarization. 
Equations (\ref{eq:gh}) and (\ref{eq:h5}) describe how the three polarizations influence the enantioselectivity via $\vec{h}$ . 

For the $E_{x+i\sigma y,x,z}$ configuration, i.e. for  $\omega$ circularly polarized in the $xy$ plane, $2\omega$ linearly polarized along $x$ and $4\omega$ linearly polarized along $z$, we have $\vec{E}_{\omega}=|E_\omega|e^{i\Phi_\omega}(\hat{x}+i\sigma\hat{y})/\sqrt{2}$, $\sigma=\pm1$, $\vec{E}_{2\omega}=|E_{2\omega}|e^{i\Phi_{2\omega}}\hat{x}$, and $\vec{E}_{4\omega}=|E_{4\omega}|e^{i\Phi_{4\omega}}\hat{z}$ which yields
\begin{equation}
\vec{h}= \frac{1}{2}i\sigma |C| e^{-i\Phi_{4\omega}}\left(\begin{array}{c}
0\\
1\\
-1
\end{array}\right),
\end{equation}
where $|C|\equiv |E_{\omega}|^2|E_{2\omega}|^2|E_{4\omega}|$, and we set the time origin so that $\Phi_{2\omega}=0$. The absence of $\Phi_{\omega}$ in the expression for $\vec{h}$ follows from the 1$\omega$ photon appearing once in both excitation pathways, which leads to its cancellation in the interference term.  

Similarly, for the $E_{x+i\sigma y,z,z}$ configuration, i.e. taking now $2\omega$ linearly polarized along $z$  ($\vec{E}_{2\omega}=|E_{2\omega}|e^{i\Phi_{2\omega}}\hat{z}$) , we obtain 
\begin{equation}
\vec{h} = i\sigma|C| e^{-i\Phi_{4\omega}}\left(\begin{array}{c}
-1\\
0\\
0
\end{array}\right).
\end{equation}
Therefore, the two configurations $E_{x+i\sigma y,x,z}$ and $E_{x+i\sigma y,z,z}$ yield  $\vec{h}$ vectors which are linearly independent of each other and thus probe complementary aspects of the molecular chirality encoded in the different components of  $\vec{g}$ [see Eq. (\ref{eq:gh})]. 

For both polarization configurations, inverting the circular polarization (reflection in the $xz$ plane) or changing the phase of $\Phi_{4\omega}$ by $\pi$  (reflection in the $xy$ plane) reverses the chirality of the field ( $\vec{h}\rightarrow-\vec{h}$). 
Since the Hamiltonian is parity invariant, inverting the chirality of the field and of the molecule simultaneously yields the same orientation-averaged population, i.e. $P^{(+)}_{R}=P^{(-)}_L$ and $P^{(+)}_{L}=P^{(-)}_R$, where $L$ and $R$  stand for $\sigma=\pm1$, respectively. Therefore, we can also express the difference between enantiomers as a difference between circular polarizations, 
\begin{equation}
P^{(+)}_{R} - P^{(-)}_{R} = P^{(-)}_{L} - P^{(-)}_{R}=-(P^{(+)}_{L} - P^{(+)}_{R}).
\label{eq:P_correspondence}
\end{equation}
Equations (\ref{eq:gh})-(\ref{eq:P_correspondence}) imply that  $P_R-P_{L}\propto \sigma\cos(\Phi_{4\omega}+\delta)$ where $\delta$ is a molecular phase difference depending on the complex nature of $\vec{g}$, which encodes all the transition dipoles and detunings associated with all photon orderings of the pathways shown in Fig. \ref{fig4}A [see Eq. (27) in Ref. \cite{ordonez2025a}] and is gauge independent.  This is precisely the nature of the oscillations observed in the SCD experiments and simulations.

\subsubsection*{2.2 Simulations in the 4-center molecular model}

\begin{figure}
	\centering
	\includegraphics[width=0.8\textwidth]{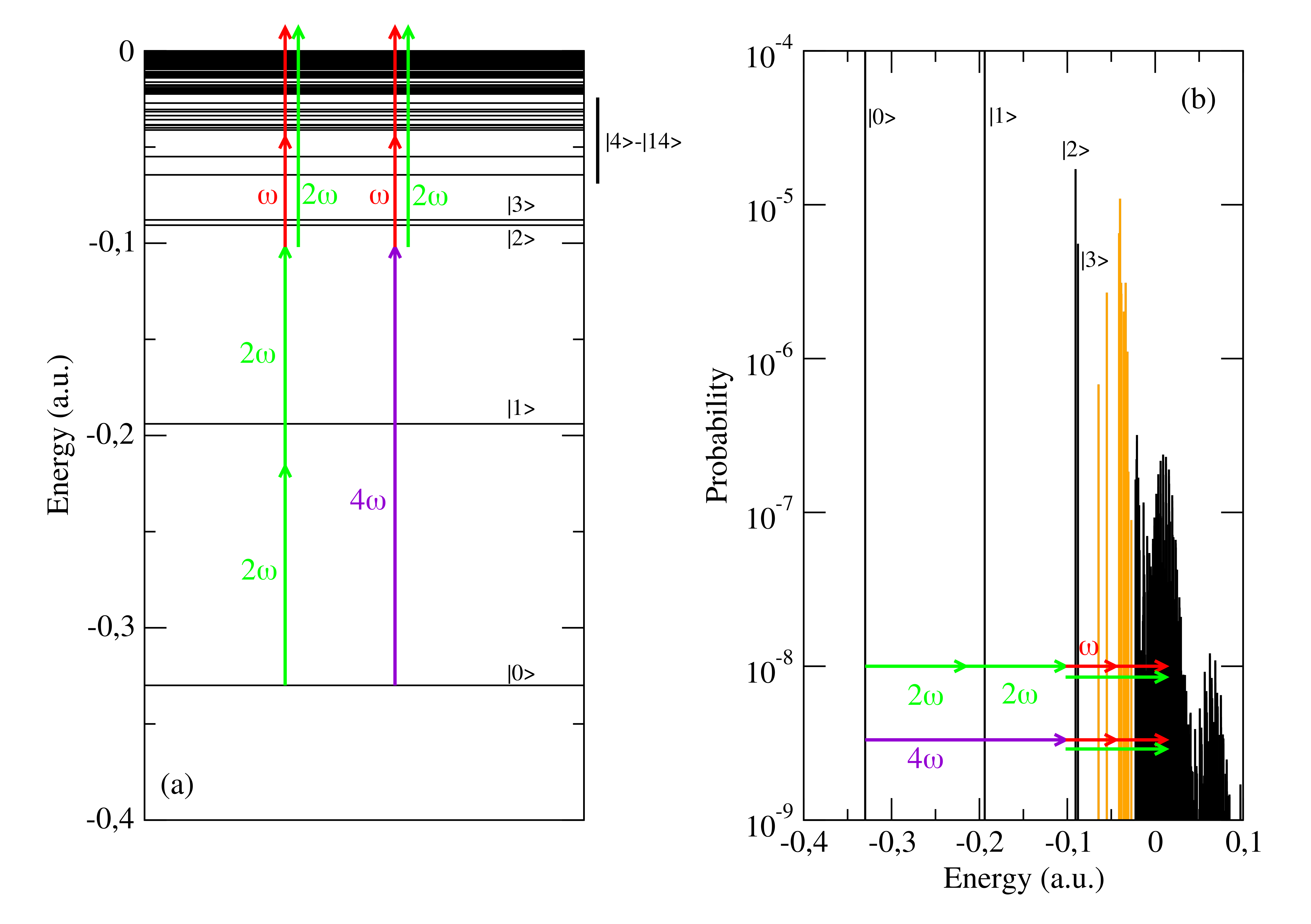} 
	\caption{
		(a) Bound energy part of the electronic spectrum of the model molecule, to which are superimposed the main excitation and ionization pathways driven by synthetic chiral light with fundamental wavelength $\lambda=800$ nm. (b) State probabilities at the end of the interaction in the $E_{x\pm iy,z,z}$ field configuration with $I_\omega=1\times10^{12}$ W.cm$^{-2}$, $I_{2\omega}=6\times10^{10}$ W.cm$^{-2}$, $I_{4\omega}=6\times10^{8}$ W.cm$^{-2}$ and $\Phi_{4\omega}=0$. The $|4-14\rangle$ states, associated to the $|0\rangle \xrightarrow{4\omega} |2,3\rangle \xrightarrow{\omega} |4-14\rangle$ and $|0\rangle \xrightarrow{2\times 2\omega} |2,3\rangle \xrightarrow{\omega} |4-14\rangle$ transitions are highlighted in orange.
        }
	\label{fig_bern_1}
\end{figure}

The chiral molecular model consists of one electron and four nuclear charges $Z_i$, with $Z_1=-1.9$ and $Z_{2-4}=0.9$, located at ${\bf R_1}={\bf 0}$, 
${\bf R_2}={\bf \hat{x}}$, ${\bf R_3}=2{\bf \hat{y}}$ and ${\bf R_4}=3{\bf \hat{z}}$. Its ionization potential is 8.98 eV which roughly coincides with the Hartree-Fock ionization potential of fenchone. The bound and continuum states $\phi_{E_i}({\bf r})$ of the system are obtained by diagonalizing the electronic Hamiltonian $H_0=-\frac{1}{2}\nabla_{\bf r}^2-\frac{Z_i}{|{\bf r}-{\bf R_i}|}$ in a set of spherical states $j_l(kr){\cal Y}_{lm}(\theta,\varphi)$, where $j_l(kr)$ are spherical Bessel functions and ${\cal Y}_{lm}(\theta,\varphi)$ are real spherical harmonics. The electronic motion is confined into a spherical box of radius $r_{max}=200$ a.u., which means that we only include the Bessel functions such that $j_l(kr_{max})=0$. The expansion is further restricted to $0 \le k \le k_{max}$ with $k_{max}=7$ a.u. and $0 \le l \le l_{max}=14$. The resulting electronic spectrum is displayed in Fig. \ref{fig_bern_1} on the energy scale restricted to its negative part -- thousands of discretized continuum states result from the diagonalization of $H_0$. 

For a given molecular orientation ${\bf \hat{R}}$, the TDSE reads in the molecular frame 
\begin{equation}
    (H_0+V({\bf \hat{R}},{\bf r}, t)-i\frac{\partial}{\partial t})\Psi({\bf \hat{R}},{\bf r},t)=0
    \label{TDSEtoy}
\end{equation}
where $\Psi({\bf \hat{R}},{\bf r},t)$ is the total electronic wavefunction and $V({\bf \hat{R}},t)$ corresponds to the laser-molecule interaction. To solve (\ref{TDSEtoy}), we expand $\Psi({\bf \hat{R}},{\bf r},t)$ onto the unperturbed $\phi_{E_i}({\bf r})$ states according to 
\begin{equation}
    \Psi({\bf \hat{R}},{\bf r},t)=\sum_i{a_i({\bf \hat{R}},t)\phi_{E_i}({\bf r})e^{-iE_it}}
    \label{expansiontoy}
\end{equation}
and we express $V({\bf \hat{R}},{\bf r},t)$ in the velocity gauge which is known to accelerate the convergence of the  
TDSE expansion \cite{cormier1996}: $V({\bf \hat{R}},{\bf r},t)=
-i{\cal R}[{\bf \hat{R}}]{\bf A}(t)\cdot{\bf \nabla_r}$ where ${\bf A}(t)$ is the vector potential associated to the 3D chiral field, which, being defined in the lab frame, is passively rotated to the molecular frame by means of ${\cal R}[{\bf \hat{R}}]$. Substituting (\ref{expansiontoy}) in (\ref{TDSEtoy}) leads to the set of coupled differential equations linking the expansion amplitudes $a_i({\bf \hat{R}},t)$
\begin{equation}
\dot{a}_i({\bf \hat{R}},t)=-{\cal R}[{\bf \hat{R}}]{\bf A}(t)\cdot\sum_j{a_j({\bf \hat{R}},t)}\langle\phi_{E_i}|\nabla_{\bf r}|\phi_{E_j}\rangle e^{-i(E_j-E_i)t}
\label{coupledtoy}
\end{equation}
which is numerically solved under the initial conditions $a_i({\bf \hat{R}},t=0)=\delta_{i,0}$ since the fundamental state is labelled with index 0. 

The orientation-averaged population of either a bound or continuum state at the end $t=\tau$ of the interaction is computed as
\begin{equation}
    (P_{E_i})^{(+,-)}_{R,L}=\int_0^{2\pi} d\alpha\int_0^\pi d\beta \sin(\beta) \int_0^{2\pi} d\gamma |a_i({\bf \hat{R}},\tau)|^2
    \label{poptoy}
\end{equation}
where $(\alpha,\beta,\gamma)$ are the Euler angles defining the orientation ${\bf \hat{R}}$. In practice, the integrals in (\ref{poptoy}) are evaluated in terms of quadratures involving Euler spacing $\Delta\alpha=\Delta\beta=\Delta\gamma=\pi/4$ radians. 

While the SCD is calculated as $2\frac{(P_{E_i})^{(+)}_{L}-(P_{E_i})^{(+)}_{R}}{(P_{E_i})^{(+)}_{L}+(P_{E_i})^{(+)}_{R}}$ for one state, XSCD for an excitation manifold is $2\frac{\sum_i[(P_{E_i})^{(+)}_{L}-(P_{E_i})^{(+)}_{R}]}{\sum_i[(P_{E_i})^{(+)}_{L}+(P_{E_i})^{(+)}_{R}]}$. SCD is computed similarly by taking into account all ionization states. 

\begin{figure}
	\centering
	\includegraphics[width=0.6\textwidth]{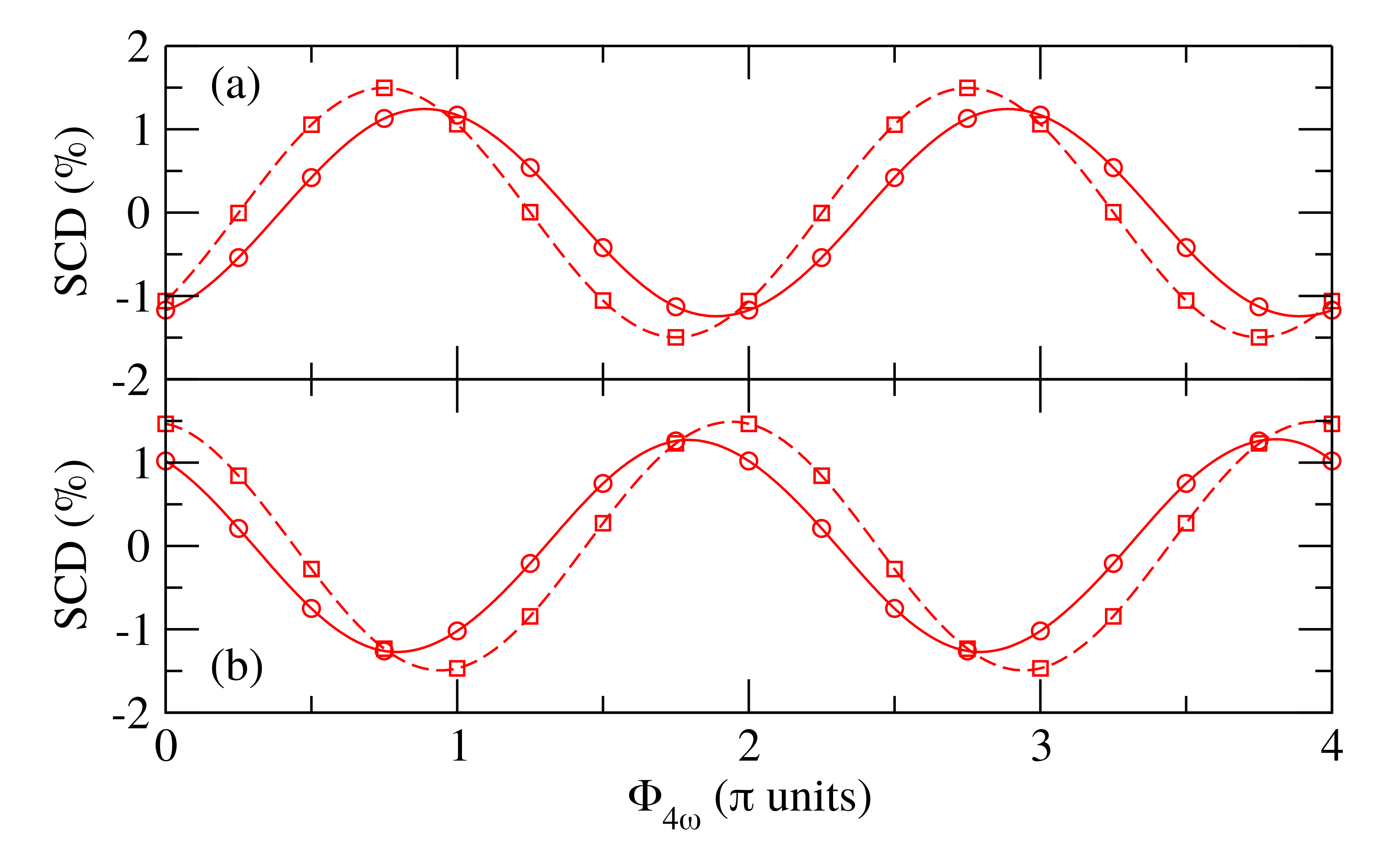} 
	\caption{
		Comparison of SCD resulting from TDSE (continuous lines with circles) and perturbative (dashed lines with squares) calculations in the 4-center model molecule, in the $E_{x\pm iy,x,z}$ (a) and $E_{x\pm iy,z,z}$ (b) field configurations with $I_\omega=1\times10^{12}$ W.cm$^{-2}$, $I_{2\omega}=6\times10^{10}$ W.cm$^{-2}$ and $I_{4\omega}=6\times10^{8}$ W.cm$^{-2}$.
        }
	\label{fig_bern_3}
\end{figure}

According to the electronic spectrum of Fig. \ref{fig_bern_1}(a), we employ a synthetic chiral field composed of a fundamental component at $\lambda=800$ nm ($1\omega=1.55$ eV), and its second and fourth harmonics. Employing the shortcut notation $|\phi_{E_i}\rangle \equiv |i\rangle$, the absorption of one photon at $4\omega$ or two photons at $2\omega$ lead to almost resonant excitation transitions $|0\rangle \rightarrow |2,3\rangle$ while additional absorption of a $1\omega$ photon leads to $|2,3\rangle \rightarrow |4-14\rangle$ transitions where $|4-14\rangle$ basically correspond to $n=3$ and the first $n=4$ Rydberg states. Threshold ionization occurs from these Rydberg states by absorption of one photon at $2\omega$, or through absorption of two $1\omega$ photons from $|2,3\rangle$ (see Fig. \ref{fig_bern_1}). State populations at the end of the interaction are exemplified in Fig. \ref{fig_bern_1}(b) in the case of $E_{x\pm iy,z,z}$ field configuration with $I_\omega=1\times10^{12}$ W.cm$^{-2}$, $I_{2\omega}=6\times10^{10}$ W.cm$^{-2}$, $I_{4\omega}=6\times10^8$ W.cm$^{-2}$ and $\Phi_{4\omega}=0$. These populations agree with the excitation and ionization pathways highlighted in Fig. \ref{fig_bern_1}(a). In order to ensure that excitation and ionization mainly proceed from these pathways, we performed perturbative calculations including $|0\rangle \xrightarrow{2\times 2\omega} |2,3\rangle \xrightarrow{\omega}|4-14\rangle$ and $|0\rangle \xrightarrow{4\omega} |2,3\rangle \xrightarrow{\omega}|4-14\rangle$ excitation pathways, complemented by the ionization channels $|2,3\rangle \xrightarrow{2\times \omega} |c\rangle $ and $|4,14 \rangle \xrightarrow{\omega}|c\rangle$ for ionization, where $|c\rangle$ refers to a continuum state. We compare in Fig. \ref{fig_bern_3} the SCD resulting from TDSE and perturbative calculations, in both $E_{x\pm iy,x,z}$ and $E_{x\pm iy,z,z}$ field configurations, for $I_\omega=1\times10^{12}$ W.cm$^{-2}$, $I_{2\omega}=6\times10^{10}$ W.cm$^{-2}$ and $I_{4\omega}=6\times10^8$ W.cm$^{-2}$. The agreement of TDSE and perturbative results is satisfactory since secondary channels such as $|0\rangle \xrightarrow{2\omega+2\times\omega} |2,3\rangle \xrightarrow{\omega}|4-14\rangle$ are not considered in the perturbative framework while they are inherently included in the infinite-order (variational) TDSE framework.

\begin{figure}
	\centering
	\includegraphics[width=0.6\textwidth]{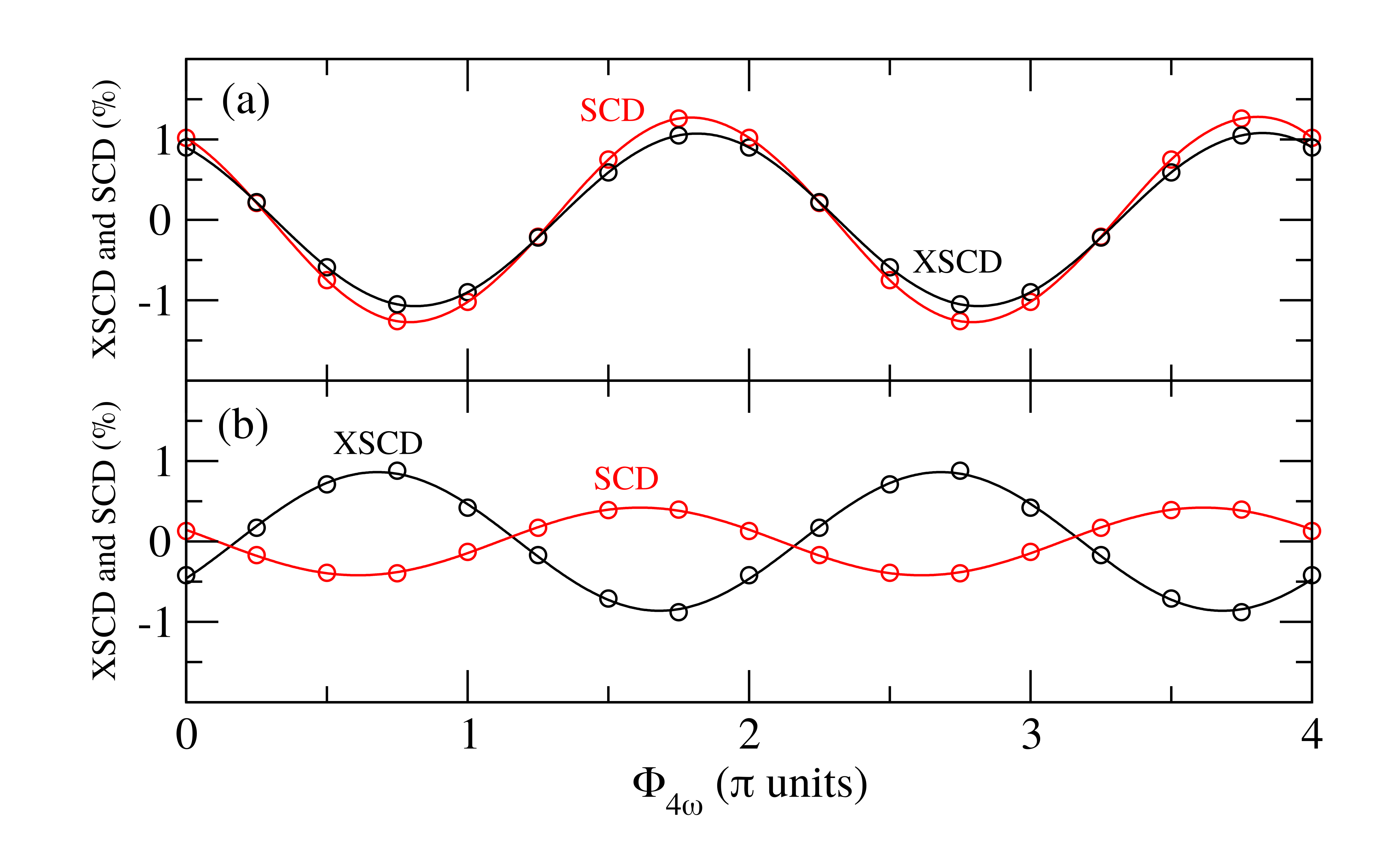} 
	\caption{
		Comparison of XSCD (black lines) and SCD (red lines) resulting from TDSE calculations in the $E_{x\pm iy,z,z}$ field configuration with $I_\omega=1\times10^{12}$ W.cm$^{-2}$ (a) and $I_\omega=4\times10^{12}$ W.cm$^{-2}$ (b). $I_{2\omega}=6\times10^{10}$ W.cm$^{-2}$ and $I_{4\omega}=6\times10^{8}$ W.cm$^{-2}$ in both panels. 
        }
	\label{fig_bern_2}
\end{figure}

We now consider the influence of the intensity $I_\omega$ of the fundamental field. We compare in Fig. \ref{fig_bern_2} the results of TDSE calculations for XSCD and SCD in the $E_{x\pm iy,z,z}$ field configuration when $I_\omega$ is increased from $1\times10^{12}$ W.cm$^{-2}$ to $4\times10^{12}$ W.cm$^{-2}$ while keeping $I_{2\omega}$ and $I_{4\omega}$ fixed to $6\times10^{10}$ W.cm$^{-2}$ and $6\times10^{8}$ W.cm$^{-2}$, respectively. We first observe that the magnitude of SCD is significantly reduced when $I_\omega$ increases. This is due to increasingly important multiphoton ionization from the fundamental state by the $1\omega$ field alone, which creates an achiral background that leads to a decrease of the SCD. The magnitude of XSCD is less influenced by $I_\omega$ but the phase of its oscillations is significantly modified. In the perturbative approach described above, the XSCD and SCD are almost independent of $I_\omega$. This is due to the fact that $|2,3\rangle \xrightarrow{2\times \omega} |c\rangle $ ionizing transitions are small compared to $|4,14 \rangle \xrightarrow{\omega}|c\rangle$ ones, so that the upper transitions, monitored by the absorption of $\omega$ photons, can be factored out in the two interaction pathways of Fig. \ref{fig_bern_1}. Therefore, the dependence of the XSCD oscillation phase on $I_\omega$ is the signature of additional interaction pathways involving absorption(s) of $\omega$ photons, beyond the channels displayed in Fig. \ref{fig_bern_1}.

\subsubsection*{2.3 Simulations in fenchone}

\begin{figure}
	\centering
	\includegraphics[width=0.6\textwidth]{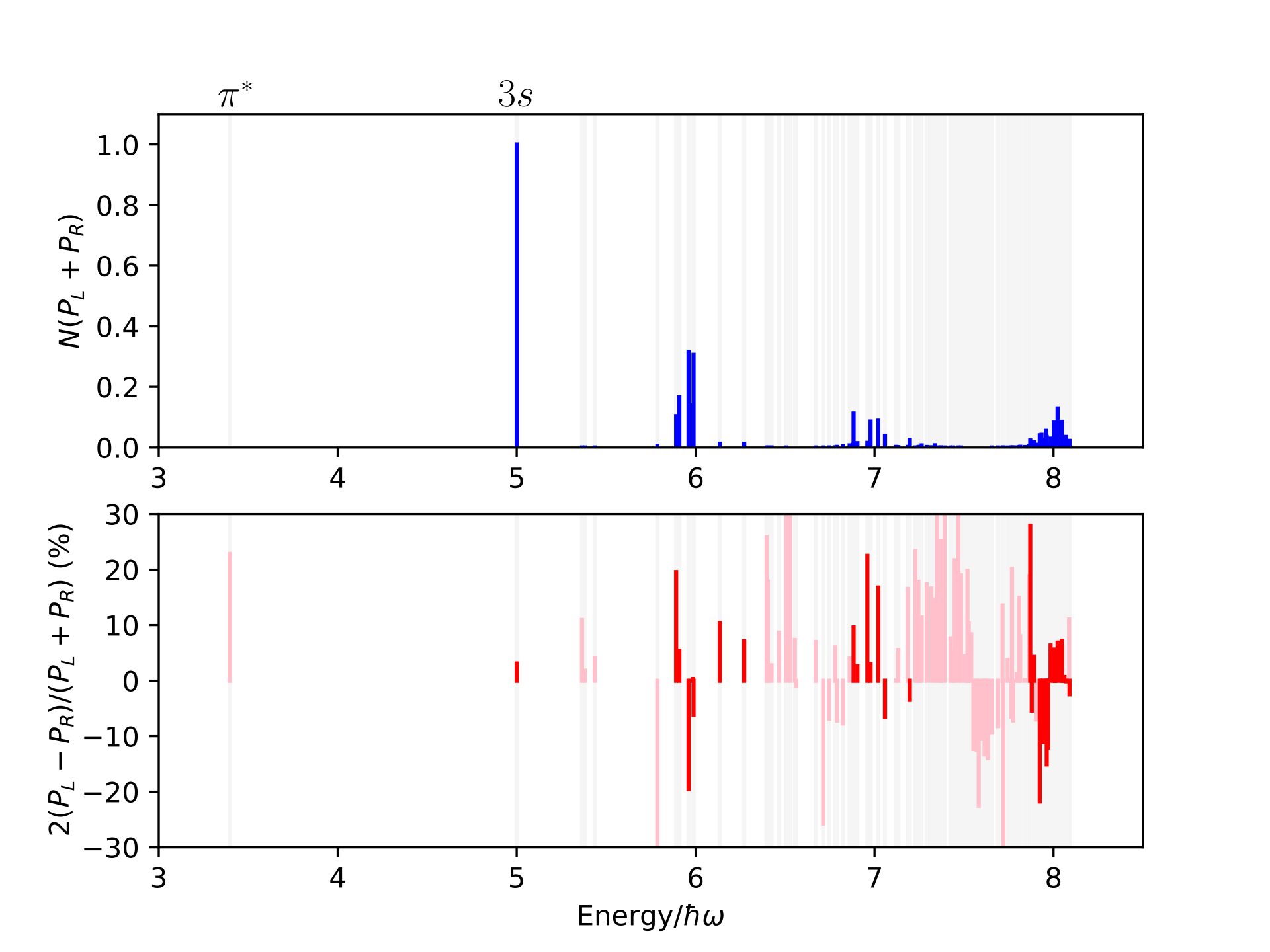} 
	\caption{
		Population and XSCD of all excited states for the $E_{x\pm iy,x,z}$ configuration in fenchone at $\Phi_{4\omega}=1.75\pi$ (see Fig. \ref{fig4}B). 
        Populations are normalized to the population of the 3s state, $N=1/[P_L(3s)+P_R(3s)]=1/(1.97\times10^{-6})$. 
        Grey lines indicate the spectral position of all excited states, dark (light) red lines indicate the XSCD for states with populations greater (smaller) than 1\% of the population of the 3s state. 
        The energy scale is with respect to the ground state and is normalized to the photon energy of the fundamental frequency. 
        }
	\label{fig:SI_populations_perp}
\end{figure}

For the simulations in fenchone molecules, electronic-structure calculations were performed to obtain the excitation energies and dipole moments required to solve the time-dependent Schrödinger equation (TDSE). The initial geometry of ($-$)-fenchone, taken from Ref. \cite{pubchem82229}, was reoptimized in this work using density functional theory (DFT) with the B3LYP functional\cite{kohndensityfunctionaltheory1996, leedevelopmentcollesalvetticorrelationenergy1988,beckedensityfunctionalexchangeenergyapproximation1988} and the 6-311G(d,p) basis set\cite{mcleancontractedgaussianbasis2008, Krishnan80}, as implemented in ORCA 6.1 \cite{ORCA6}. Excitation energies, permanent  and transition dipole moments for the first 100 excited states were then calculated using time-dependent density functional theory (TDDFT) \cite{TDDFT} with the full random-phase approximation (RPA), using the CAM-B3LYP functional \cite{YANAI200451} and the aug-cc-pVDZ basis set\cite{dunning89,augccpvdz}, as implemented in Q-Chem 5.0 \cite{QChem}.
The TDSE was solved as explained in the previous section, but in the length gauge where the interaction term reads $V({\bf \hat{R}},\{ {\bf r}_i \},t)={\cal R}[{\bf \hat{R}}]{\bf E}(t)\cdot\sum_i{\bf r}_i$, with ${\bf r}_i$ the electron coordinate of the $i$-th electron in fenchone. We further employed Gaussian pulses of 20.5 fs (FWHM of intensity) for the three frequencies $\omega$, $2\omega$, and $4\omega$. 
The value of $\omega$ was chosen so that $5\omega=6.29$ eV is resonant with the transition from the ground state to the 3s state. 
The final excited state populations were averaged over 2380 orientations using a Lebedev grid of order 21 (170 points) for two of the Euler angles and an equally spaced grid with 14 points for the third Euler angle. 
We verified that the population of the 3s state satisfies the expected symmetries $P^{(+)}_{R}=P^{(-)}_L$, $P^{(-)}_{R}=P^{(+)}_L$, $P(\Phi_\omega)=P(\Phi_\omega+\Delta\Phi_\omega)$ for arbitrary $\Delta\Phi_\omega$ up to negligibly small numerical variations. 
Fig. \ref{fig:SI_populations_perp} shows the populations (top) and XSCD (bottom) for all excited states at the value of $\Phi_{4\omega}=1.75\pi$ (see Fig. \ref{fig4}B). 





\end{document}